\documentclass[11pt,a4paper]{article}
\pdfoutput=1

\newif\iffigs\figstrue

\usepackage{jheppub}
\usepackage[utf8]{inputenc}
\usepackage[T1]{fontenc}
\usepackage{epsfig,latexsym}
\usepackage{amsmath}
\usepackage{xcolor}
\usepackage{adjustbox}
\usepackage{graphicx}
\usepackage{verbatim}
\usepackage{mathrsfs}
\usepackage{amssymb}
\usepackage{multirow}
\usepackage{epsfig}
\usepackage{color,colordvi}
\usepackage{appendix}
\usepackage{slashed}
\usepackage{cancel}
\usepackage{float}
\usepackage[symbol]{footmisc}
\usepackage[justification=centering]{caption}
\usepackage{epsf}
\usepackage{amsmath}
\usepackage{physics}
\usepackage{MnSymbol}

 \csname
@addtoreset\endcsname{equation}{section}
\def\be{\begin{equation}}
\def\ee{\end{equation}}
\def\bes{\begin{equation*}}
\def\ees{\end{equation*}}
\def\bead{\begin{aligned}}
\def\eead{\end{aligned}}
\def\bmat{\left(\begin{matrix}}
\def\emat{\end{matrix}\right)}
\def\cL{{\cal L}}
\def\cC{{\cal C}}

\def\cO{{\cal O}}

\vspace{0.5cm}

\hypersetup{pdfstartview=FitV,colorlinks=true,linkcolor=blue,citecolor=red,filecolor=black,urlcolor=blue}

\title{\boldmath Chiral Froggatt-Nielsen models, gauge anomalies and flavourful axions}
\author[a,b]{Q. Bonnefoy}
\author[b]{, E. Dudas}
\author[c]{and S. Pokorski}

\affiliation[a]{DESY, Notkestrasse 85, 22607 Hamburg, Germany}
\affiliation[b]{Centre de Physique Th{\'e}orique, CNRS, {\'E}cole Polytechnique, IP Paris, 91128 Palaiseau, France}
\affiliation[c]{Institute of Theoretical Physics, Faculty of Physics, University of Warsaw, ul. Pasteura 5, PL-02-093 Warsaw, Poland}

\emailAdd{quentin.bonnefoy@desy.de}
\emailAdd{emilian.dudas@polytechnique.edu}
\emailAdd{stefan.pokorski@fuw.edu.pl}

\abstract{We study UV-complete Froggatt-Nielsen-like models for the generation of mass and mixing hierarchies, assuming that the integrated heavy fields are chiral with respect to an abelian Froggatt-Nielsen symmetry. It modifies the mixed anomalies with respect to the Standard Model gauge group, which opens up the possibility to gauge the Froggatt-Nielsen symmetry without the need to introduce additional spectator fermions, while keeping mass matrices usually associated to anomalous flavour symmetries. We give specific examples where this happens, and we study the flavourful axion which arises from an accidental Peccei-Quinn symmetry in some of those models. Such an axion is typically more coupled to matter than in models with spectator fermions.}

\preprint{{\raggedleft DESY 19-154\\CPHT-RR051.082019 \par}}


\begin{document} 
\maketitle
\flushbottom

\section{Introduction}

As efficient as the Standard Model (SM) may be to describe particle physics phenomenology, it still has unsatisfactory features. Among those, the unexplained hierarchies in masses and mixings between elementary particles has motivated intense theoretical work, leading to precise BSM scenarii. The latter deal with the flavour hierarchies, as well as with the several discrepancies with the SM predictions in magnetic dipole moments or heavy meson decays, while abiding by the conclusions of precision tests of the SM.

Many flavour models for the mass hierarchies involve additional symmetries, whose nature and origin are diverse. In particular, Froggatt-Nielsen (FN) models \cite{Froggatt:1978nt} (see also \cite{Dimopoulos:1983rz,Leurer:1992wg,Leurer:1993gy} and references therein) are leading candidates to account for the flavour hierarchies. They rely on an extended scalar and fermionic heavy sector and on an additional spontaneously broken symmetry. Their study has recently been revived by the focus on flavourful axions which arise in FN-like setups \cite{Davidson:1981zd,Wilczek:1982rv,Ema:2016ops,Calibbi:2016hwq,Ema:2018abj} (see also \cite{Davidson:1983fy,Davidson:1983tp,Davidson:1984ik,Geng:1988nc,Berezhiani:1989fp,Berezhiani:1990wn,Berezhiani:1990jj,Babu:1992cu,Sakharov:1994pr,Albrecht:2010xh,Cheung:2010hk,Ahn:2014gva,Celis:2014iua,Nomura:2016nfi,Arias-Aragon:2017eww,Bjorkeroth:2017tsz,Ahn:2018nfb,Ahn:2018cau,Linster:2018avp,Suematsu:2018hbu,Bjorkeroth:2018ipq} for other studies of flavourful axions) and whose EFT is very much constrained by flavour physics \cite{Choi:2017gpf,Bjorkeroth:2018dzu,Gavela:2019wzg}. Such flavourful axions can also be linked with dark matter studies \cite{Jaeckel:2013uva}.

The nature of the FN symmetry is debatable, and the question of whether it can be gauged is raised, in particular in order to evade quantum gravity corrections which explicitly break global symmetries \cite{Hawking:1987mz,Giddings:1988cx,Banks:2010zn,Harlow:2018jwu,Harlow:2018tng,Fichet:2019ugl}. For instance, it has been shown \cite{Ibanez:1994ig,Jain:1994hd,Binetruy:1994ru,Dudas:1995yu} that in minimal supersymmetric models, the MSSM spectrum induces gauge anomalies when charged under a FN symmetry, such that one must design an extra fermionic spectrum or a Green-Schwarz (GS) mechanism in order to make the model consistent (see e.g. \cite{Dudas:1996fe,Binetruy:1996xk,Irges:1998ax,Froggatt:1998he,King:1999mb,King:1999cm,Berger:2000sc,Dreiner:2003yr,Chen:2006hn,Dreiner:2007vp,DelleRose:2017xil,Bonnefoy:2018hdo}). One way to do this is to add chiral spectator fermions at the scale where the FN symmetry is broken.

In this paper, we explore the possibility of gauging the FN symmetry without adding any other extra field than the ones required to implement the FN mechanism. In particular, we do not need to introduce both heavy vector-like fields which generate the flavour hierarchies {\`a la} Froggatt-Nielsen and chiral ones which take care of anomalies. This is indeed possible if the fields participating in the FN mechanism are chiral with respect to the FN symmetry, which we choose to be abelian in what follows. We show it by presenting specific examples of two kinds, without (as already shown in \cite{Alonso:2018bcg,Smolkovic:2019jow}) and with  an accidental global symmetry, focusing for concreteness on supersymmetric models (we briefly comment on non-SUSY models at the end of the discussion).
In particular, we explicitly display a model with a physical flavourful axion, which we analyze and compare to flavourful axions arising from global FN symmetries, dubbed flaxions/axiflavons \cite{Ema:2016ops,Calibbi:2016hwq,Ema:2018abj}. In our example, although the qualitative axion phenomenology is similar to the one of global flaxions/axiflavons, meaning that the axion couplings are mainly dictated by low-energy physics, there are slight changes in the axion couplings to gauge fields since the latter are already generated by the integrating-out of the heavy FN sector. An other obvious but significant difference between the global and the gauged FN models with an axion is that the shift symmetry of the latter can easily be protected in the second kind of models.

We also establish constraints coming from the perturbativity of the (MS)SM gauge couplings, which imposes that the scale of spontaneous FN symmetry breaking is at least intermediate ($10^{12-13}$ GeV). This allows us to compare chiral FN models with gauged vector-like FN models which use spectator fields to cancel the anomalies. An immediate consequence of using a chiral heavy sector instead of a vector-like one is that there are in general less SM-charged heavy particles, such that constraints from the running of gauge couplings are weaker. In particular, axions can be more coupled to matter in chiral models. An other feature of the latter is that they reduce the number of necessary input scales, since they do not need to introduce the mass scale of the vector-like FN fields.

The plan of the paper is as follows: in section \ref{motivSection}, we review minimal supersymmetric abelian FN models and their naive gauging to motivate the present work. In section \ref{sectionMain}, we study in details how the conclusions of section \ref{motivSection} are evaded if the fields which generate the mass and mixing hierarchies are chiral with respect to the FN symmetry. We discuss our general framework in section \ref{chiralGeneral}, illustrate it with specific examples in section \ref{chiralExamples}, while section \ref{gaugeRunningSection} presents the constraints coming from the running of the MSSM gauge couplings. At this point, we also compare chiral and vector-like gauged FN models, in the spirit of what was sketched above. In section \ref{accidentalAxion}, we elaborate on the accidental flavourful Peccei-Quinn symmetry \cite{Peccei:1977hh,Weinberg:1977ma,Wilczek:1977pj} and its associated axion which arise in some of the models we scrutinize, and we briefly discuss constraints on the model parameters derived from the consistency of the model. We also discuss constraints arising from the axion phenomenology in section \ref{axionPhenoSession}. We mention non-SUSY models in section \ref{nonSUSY}, by discussing again some examples. After a conclusive summary, appendix \ref{fermionCouplingsAppendix} covers a discussion of gauge-invariant superpotential terms which were postponed in section \ref{chiralGeneral}, appendix \ref{anomaliesUnificationAppendix} discusses the link between the GS conditions for anomaly cancellation and the unification of gauge couplings and appendix \ref{vectorlikeAppendix} details the construction of some of the gauged vector-like FN models discussed in section \ref{gaugeRunningSection}. The preliminary results of this work have been presented in \cite{DudasPlanck2019}.

\section{Motivation}\label{motivSection}

\subsection{Yukawa matrices and Froggatt-Nielsen models}

The flavour structure of the SM is a consequence of the number of families and of the structure of the Yukawa sector:
\be
\cL \supset -(Y^u_{ji}\overline{u_{R,i}}HQ_{L,j}+Y^d_{ji}\overline{d_{R,i}}H^cQ_{L,j}+Y^e_{ji}\overline{e_{R,i}}H^cL_{L,j})+h.c. \ .
\label{SMYukawas}
\ee
Its phenomenological predictions are fully characterized by fermion masses $m_{i=1..3}^X$ (with $X=u,d,e$, a notation we use throughout this paper) and by the CKM matrix \cite{Cabibbo:1963yz,Kobayashi:1973fv}. The latter reads, in the Wolfenstein parametrization \cite{Wolfenstein:1983yz}:
\be
V_\text{CKM}=\bmat1-\frac{\lambda^2}{2}&\lambda&A\lambda^3(\rho-i\eta)\\-\lambda&1-\frac{\lambda^2}{2}&A\lambda^2\\A\lambda^3(1-\rho-i\eta)&-A\lambda^2&1\emat+\cO(\lambda^4) \ ,
\label{hierarchiesMixings}
\ee
where $\lambda$ is linked to the Cabibbo angle $\theta_C$: $\lambda=\sin(\theta_C)\approx 0.22$, and $A,\rho,\eta=\cO(1)$. Orders of magnitude for the quark and lepton masses can also be expressed in terms of $\lambda$:
\be
\bead
&\frac{m_u}{m_t}\sim\lambda^8 \ , \ \frac{m_c}{m_t}\sim\lambda^4 \ , \ \frac{m_d}{m_b}\sim\lambda^4 \ , \ \frac{m_s}{m_b}\sim\lambda^2 \ , \ \frac{m_b}{m_t}\sim\lambda^2\ ,\ \frac{m_e}{m_\tau}\sim\lambda^4\ , \ \frac{m_\mu}{m_\tau}\sim\lambda^2\ , \ \frac{m_\tau}{m_t}\sim\lambda^2
\eead
\label{hierarchiesMasses}
\ee
at the GUT scale $M_\text{GUT}\sim 10^{16}$ GeV. The strong hierarchies between the particle masses as well as the milder ones appearing in the CKM matrix are unexplained input parameters in the SM. They can be traced back to hierarchies which must be present in the Yukawa matrices $Y^{u,d,e}$ of \eqref{SMYukawas}. In the minimal supersymmetric standard model (MSSM), to which we stick except in section \ref{nonSUSY}, the flavour structure as well as the mass and mixing hierarchies are found in the superpotential\footnote{Our definitions and conventions for the MSSM superfields can be found in Table \ref{MSSMcharges}.}:
\be
W \supset Y^u_{ij}Q_{i}H_uU_{j}+Y^d_{ij}Q_{i}H_dD_{j}+Y^e_{ij}L_{i}H_dE_{j} \ .
\label{MSSMYukawas}
\ee

Froggatt-Nielsen models \cite{Froggatt:1978nt} address the origin of flavour hierarchies by means of a symmetry explanation: the masses and mixings arise after spontaneous breaking of a chiral symmetry, which forbids their existence when it is exact in the UV (except for the top quark Yukawa term, as well as the bottom quark one if $\tan\beta$ is large). For instance, one can postulate a global horizontal/family symmetry $U(1)_\text{FN}$ acting on the different MSSM fields and on a standard model singlet superfield $\phi$, the flavon. Then, $U(1)_\text{FN}$ invariance of the Yukawa sector of the MSSM requires a dressing of the Yukawa matrices by powers of $\phi$:
\be
W \supset h^u_{ij} \left(\frac{\phi}{M}\right)^{n^u_{ij}}Q_{i}H_uU_{j}+h^d_{ij} \left(\frac{\phi}{M}\right)^{n^d_{ij}}Q_{i}H_dD_{j}+
h^e_{ij} \left(\frac{\phi}{M}\right)^{n^e_{ij}}L_{i}H_dE_{j} \ ,
\label{minimalFN}
\ee
where the $h^X_{ij}$ are order one numbers, $M$ is a high scale of new physics, for instance the mass scale of heavy fields which mix with the standard model ones (see section \ref{chiralGeneral} for explicit examples) or the Planck mass if those higher-dimensional operators are generated by supergravity, and the $n^X_{ij}$ are the $U(1)_\text{FN}$ charges of the MSSM Yukawa couplings in units of the charge of $\overline\phi$. Indeed, $U(1)_\text{FN}$ invariance imposes that the $n^X$'s are
\be
n^u_{ij}=-\frac{q_{Q_i}+q_{U_j}+q_{H_u}}{q_\phi} \ , \quad n^d_{ij}=-\frac{q_{Q_i}+q_{D_j}+q_{H_d}}{q_\phi} \ , \quad n^e_{ij}=-\frac{q_{L_i}+q_{E_j}+q_{H_d}}{q_\phi} \ ,
\label{linkNQ}
\ee
where the $q$'s denote with transparent subscripts the $U(1)_\text{FN}$ charges of the different superfields. In particular, they are
such that
\be
n^X_{11}-n^X_{i1}=n^X_{1j}-n^X_{ij} \ , \quad n^u_{11}-n^u_{i1}=n^d_{11}-n^d_{i1} \ .
\ee
Once $U(1)_\text{FN}$ is spontaneously broken by a vacuum expectation value (vev) of $\phi$, the hierarchies in the fermion mass matrices are naturally explained in terms of a small parameter $\epsilon=\abs{\frac{\langle\phi\rangle}{M}}$, assumed to be $\sim\lambda$, and larger charges for the light generations (see section \ref{chiralExamples} for explicit examples). Indeed,  the low-energy Yukawa couplings are given by
\begin{equation}
Y_{ij}^X =  h_{ij}^X \epsilon^{n_{ij}^X}      
\end{equation}
and have the required hierarchies for $h_{ij}^X \sim \cO(1)$.

\subsection{Gauged $U(1)_\text{FN}$ and anomaly cancellation}\label{anomalousFN}

Since global symmetries are threatened by quantum gravity \cite{Hawking:1987mz,Giddings:1988cx,Banks:2010zn,Harlow:2018jwu,Harlow:2018tng,Fichet:2019ugl}, one could be tempted to gauge $U(1)_\text{FN}$ to protect it against explicit breaking, which could in principle generate uncontrolled $U(1)_\text{FN}$-breaking Yukawa terms and spoil the symmetry-based hierarchies \eqref{hierarchiesMixings}-\eqref{hierarchiesMasses}. However, the possible $U(1)_\text{FN}$ charges are constrained by the flavour structure of the SM and the question of whether they can be chosen such that all gauge anomalies vanish is raised \cite{Ibanez:1994ig,Jain:1994hd,Binetruy:1994ru,Dudas:1995yu}. In particular, defining anomaly coefficients such that 
\be
\bead
\delta_{U(1)_\text{FN}}\cL=&-\frac{A_3}{192\pi^2}\epsilon^{\mu\nu\rho\sigma}G_{\mu\nu}^aG_{\rho\sigma}^a-\frac{A_2}{192\pi^2}\epsilon^{\mu\nu\rho\sigma}W_{\mu\nu}^iW_{\rho\sigma}^i-\frac{A_1}{192\pi^2}\epsilon^{\mu\nu\rho\sigma}F_{\mu\nu}F_{\rho\sigma}-...\\
&+\cO(\text{three gauge boson terms}) \ ,
\eead
\ee
where $G_\mu^a$ is the gluon field of field strength $G_{\mu\nu}^a$, $W_{\mu(\nu)}^i$ the $SU(2)_W$ gauge boson field (strength) and $F_{\mu(\nu)}$ the $U(1)_Y$ gauge boson field (strength), it has been shown that\footnote{Many order one coefficients have been and will be dropped, which may change slightly the estimated orders of magnitude. However, they do not change qualitatively the anomaly discussion.}
\be
\abs{\det(Y_uY_d^{-2}Y_e^3)}\sim\epsilon^{\frac{3}{2}(A_1+A_2-2A_3)} \ .
\ee
The determinant of the left hand side is clearly $\epsilon$-suppressed when we insert phenomenologically relevant Yukawa matrices. For instance, assuming $\tan\beta=1$,
\be
\abs{\det(Y_uY_d^{-2}Y_e^3)}\sim\frac{m_um_cm_tm_e^3m_\mu^3m_\tau^3}{v^6m_d^2m_s^2m_b^2}\sim \lambda^{30} \ ,
\label{YuYdfirst}
\ee
where $v$ is the Higgs vev. Thus, we understand that \eqref{minimalFN} introduces a $U(1)_\text{FN}$ which has mixed anomalies with the SM gauge group $G_\text{SM}$. This enables one to interpret the phase of $\phi$ as a flavourful QCD axion when $U(1)_\text{FN}$ is global \cite{Wilczek:1982rv,Ema:2016ops,Calibbi:2016hwq,Ema:2018abj}, with couplings to gauge fields and SM fermions which are fully determined by the mass matrices. On the other hand, it also means that $U(1)_\text{FN}$ cannot be naively gauged. Ways out would either introduce additional chiral fermions, extend the scalar sector or rely on a Green-Schwarz-inspired mechanism \cite{Green:1984sg}. In what follows, we will explore the first and second options. Let us point out in passing that, due to the phenomenological interest in additional abelian factors to the SM gauge group, there are recent works about anomaly cancellation in such models with a general focus, see e.g. \cite{Ellis:2017nrp,Allanach:2018vjg,Correia:2019pnn,Costa:2019zzy}.

\section{Chiral Froggatt-Nielsen models}\label{sectionMain}

A key assumption in \eqref{minimalFN}, which we now relax, is that the only low-energy contribution of the heavy sector at scale $M$ is the generation of the Yukawa terms. This is true if the heavy sector is vector-like with respect to the SM gauge group, but if it is chiral there could also be in the EFT anomalous couplings between (the longitudinal component of) the $U(1)_\text{FN}$ gauge field and the SM gauge bosons \cite{Anastasopoulos:2006cz}. In this section, we explore this possibility. 

We focus on models with two singlet superfields $\phi_1$ and $\phi_2$, which respectively replace the flavon $\phi$ and the mass $M$ in \eqref{minimalFN}, such that the Yukawa sector is as follows:
\be
W \supset h^u_{ij} \left(\frac{\phi_1}{\phi_2}\right)^{n^u_{ij}}Q_{i}H_uU_{j}+
h^d_{ij} \left(\frac{\phi_1}{\phi_2}\right)^{n^d_{ij}}Q_{i}H_dD_{j}+
h^e_{ij} \left(\frac{\phi_1}{\phi_2}\right)^{n^e_{ij}}L_{i}H_dE_{j} \ ,
\label{chiralFN}
\ee
and we allow in particular $\phi_2$ to be charged under $U(1)_\text{FN}$. Generalizing \eqref{linkNQ}, we now have
\be
n^u_{ij}=-\frac{q_{Q_i}+q_{U_j}+q_{H_u}}{q_{\phi_1}-q_{\phi_2}} \ , \quad n^d_{ij}=-\frac{q_{Q_i}+q_{D_j}+q_{H_d}}{q_{\phi_1}-q_{\phi_2}} \ , \quad n^e_{ij}=-\frac{q_{L_i}+q_{E_j}+q_{H_d}}{q_{\phi_1}-q_{\phi_2}} \ ,
\label{linkNQChiral}
\ee
and we define $x_{1,2}\equiv -q_{\phi_{1,2}},h_{u,d}\equiv q_{H_{u,d}}$ for the sake of reducing the subscripts in what follows. In order to trace back the role of the hierarchies in the Yukawa matrices, we also trade most of the charges for the integers $n^X_{ij}$ using \eqref{linkNQChiral}, such that for instance
\be
q_{U_1}=-q_{Q_1}-h_u+(x_1-x_2)n^u_{11} \ .
\ee
Working out other relations leads to the charges of the superfields which appear in Table \ref{MSSMcharges}.
\begin{table}[h]
\bes
\begin{array}{|c|c|c|c|c|}
\hline
&SU(3)_C&SU(2)_W&U(1)_Y&U(1)_\text{FN}\\
\hline
\phi_1&\textbf{1}&\textbf{1}&0&-x_1\\
\phi_2&\textbf{1}&\textbf{1}&0&-x_2\\
H_u&\textbf{1}&\textbf{2}&1/2&h_u\\
H_d&\textbf{1}&\textbf{2}&-1/2&h_d\\
Q_{i}&\textbf{3}&\textbf{2}&1/6&X_Q-(x_1-x_2)(n^u_{11}-n^u_{i1})\\
U_{j}&\overline{\textbf{3}}&\textbf{1}&-2/3&-X_Q-h_u+(x_1-x_2)n^u_{1j}\\
D_{j}&\overline{\textbf{3}}&\textbf{1}&1/3&-X_Q-h_d+(x_1-x_2)n^d_{1j}\\
L_{i}&\textbf{1}&\textbf{2}&-1/2&X_L-(x_1-x_2)(n^e_{11}-n^e_{i1})\\
E_{j}&\textbf{1}&\textbf{1}&1&-X_L-h_d+(x_1-x_2)n^e_{1j}\\
\hline
\end{array}
\ees
\caption{Gauge charges of the singlet and MSSM fields\\$X_Q\equiv q_{Q_1}$ is the $U(1)_\text{FN}$ charge of $Q_{1}$, $X_L\equiv q_{L_1}$ the $U(1)_\text{FN}$ charge of $L_{1}$}
\label{MSSMcharges}
\end{table}
We again assume $\langle\phi_1\rangle=\epsilon\langle\phi_2\rangle$, with $\epsilon\approx\lambda$. Even though the thorough discussion of the scalar potential of the theory is beyond the scope of this paper, we note that obtaining suitable vevs out of it seems to be possible\footnote{One may wonder whether the presence of the $U(1)_\text{FN}$ group  constrains  the system so much that $\langle\phi_1\rangle=\epsilon\langle\phi_2\rangle\sim$ (a target scale) is hard to ensure. We believe that this is not the case. Indeed, 
keeping all MSSM mass scales - including the soft terms - close to the weak scale, the vevs of $\phi_1$ and $\phi_2$ can be (approximately) obtained from  the potential restricted to them alone. In this potential, one finds the $U(1)_\text{FN}$ D-term as well as possible F-terms and SUGRA corrections. This brings already quite some freedom to get the right vevs: for instance, if $x_1$ and $x_2$ have the same sign there is no F-term, and SUGRA corrections can come from K\"ahler terms such as $\abs{X}^2\phi_1^{x_2}\overline{\phi_2}^{x_1}+h.c.$, where $X$ is a chiral superfield from the SUSY-breaking sector. The (restricted) scalar potential is
$
V_{\phi_1,\phi_2}=V_D+m_{3/2}\phi_1^{x_2} \overline {\phi_2}^{x_1} +h.c.
$
(with $M_P=1$) and minimization brings $x_1^2 |\langle\phi_1\rangle|^2 =  x_2^2 |\langle\phi_2\rangle|^2$. Adding a Fayet-Iliopoulos term $\xi$ in the D-term, we also get $\langle\phi_{1,2}\rangle\sim\xi$. Without SUSY as in section \ref{nonSUSY}, there is enough freedom in the potential to engineer vevs at desired values.}. Formulas to follow will encompass cases where $\phi_1$ or $\phi_2$ is uncharged and equivalent to a mass $M$, but we always impose $x_1\neq x_2$ such that $U(1)_\text{FN}$ acts non-trivially on the MSSM fields.

The contribution of the MSSM fields to the mixed anomaly coefficients are as follows:
\be
\bead
SU(3)_C^2\times U(1)_\text{FN}: \ A_{3,\text{SM}} =& \ \sum_i(2q_{Q_i}+q_{U_i}+q_{D_i})= -3(h_u+h_d)+(x_1-x_2)\sum_i(n^u_{ii}+n^d_{ii})\\
SU(2)_W^2\times U(1)_\text{FN}: \ A_{2,\text{SM}}=& \ \sum_i(3q_{Q_i}+q_{L_i})+q_{H_u}+q_{H_d}\\
=& \ 3(3X_Q+X_L)+h_u+h_d\\
&-(x_1-x_2)\Big(3(2n^u_{11}-n^u_{21}-n^u_{31})+2n^e_{11}-n^e_{21}-n^e_{31}\Big)\\
U(1)_Y^2\times U(1)_\text{FN}: \ A_{1,\text{SM}}=& \ \sum_i\left(\frac{q_{Q_i}}{3}+\frac{8q_{U_i}}{3}+\frac{2q_{D_i}}{3}+q_{L_i}+2q_{E_i}\right)+q_{H_u}+q_{H_d}\\
=& -3(3X_Q+X_L)-7(h_u+h_d)\\
&+(x_1 - x_2)\Bigg(\frac{n^u_{21}+n^u_{31}-2n^u_{11}}{3}+\frac{8(n^u_{11}+n^u_{12}+n^u_{13})}{3}\\
&\qquad\qquad\qquad+\frac{2(n^d_{11}+n^d_{12}+n^d_{13})}{3}+2n^e_{12}+2n^e_{13}+n^e_{21}+n^e_{31}\Bigg) \ .
\eead
\label{newSManomalies}
\ee
The vanishing of the mixed $U(1)_Y\times U(1)_\text{FN}^2$ anomaly is also imposed, but for brevity we do not display it explicitly. On the other hand, we ignore the $U(1)_\text{FN}^3$ or $U(1)_\text{FN}\times$gravity anomalies. Those could for instance be modified if we added to this setup some sterile neutrino superfields charged under $U(1)_\text{FN}$. 

The anomalies in \eqref{newSManomalies} are non-vanishing, since the discussion of section \ref{anomalousFN} still applies. Nonetheless, they can be cancelled by taking into account the gauge anomalies induced by the heavy FN sector, as we now discuss.

\subsection{Heavy FN sector and anomalies}\label{chiralGeneral}

\begin{figure}
\centering
\includegraphics[scale=0.5]{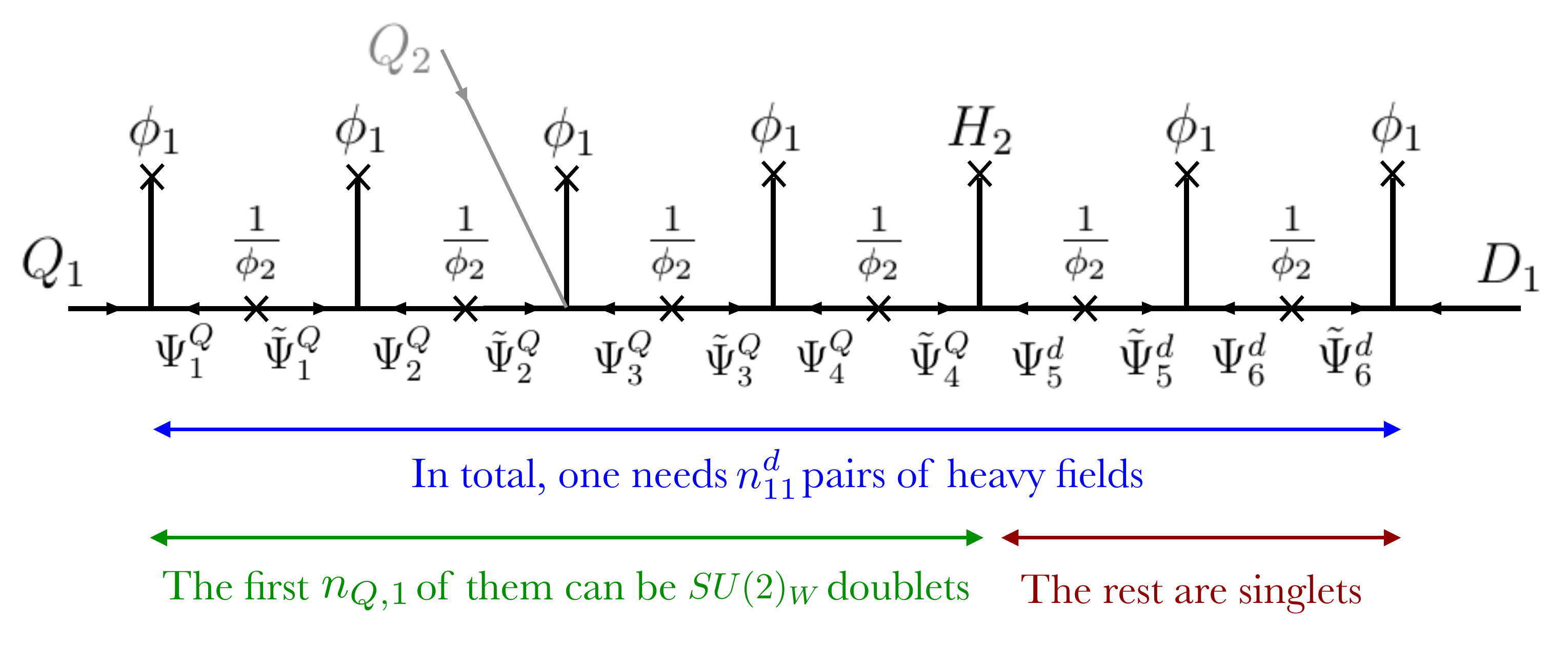}
\caption{Tree diagram generating the $d$-quark mass, when $n^d_{11}=6$ and $n_{Q,1}=4$\\
The gray line indicates how it should be modified to generate a mixing to $Q_{2}$ when $n^d_{21}=4$}
\label{treeDiags}
\end{figure}

We now design a UV theory which generates \eqref{chiralFN} in the IR. We understand \eqref{chiralFN} as being perturbatively generated\footnote{All the operators we consider in this paper are either perturbatively generated or present in the UV.}, closely following the original FN picture. The setup, together with our notations, can be understood by looking at Figure \ref{treeDiags}: ignoring for the time being the second and third SM generations, we introduce the heavy fermions shown in Table \ref{Heavycharges}, vector-like under the SM gauge group but chiral with respect to $U(1)_\text{FN}$.
\begin{table}[h]
\bes
\begin{array}{|c|c|c|c|c|}
\hline
&SU(3)_C&SU(2)_W&U(1)_Y&U(1)_\text{FN}\\
\hline
\multicolumn{5}{|c|}{\text{Common to all quarks}}\\
\hline
\Psi^Q_{i=1,...,n_{Q,1}\leq\max[n^u_{11},n^d_{11}]}&\overline{\textbf{3}}&\textbf{2}&-1/6&-X_Q+i(x_1-x_2)+x_2\\
\tilde\Psi^Q_{i=1,...,n_{Q,1}\leq \max[n^u_{11},n^d_{11}]}&\textbf{3}&\textbf{2}&1/6&X_Q-i(x_1-x_2)\\
\hline
\multicolumn{5}{|c|}{\text{For $U$'s or $D$'s}}\\
\hline
\Psi^u_{i=n_{Q,1}+1,...,n^u_{11}}&\overline{\textbf{3}}&\textbf{1}&-2/3&-X_Q+(i-1)(x_1-x_2)-h_u\\
\tilde\Psi^u_{i=n_{Q,1}+1,...,,n^u_{11}}&\textbf{3}&\textbf{1}&2/3&X_Q-(i-1)(x_1-x_2)+x_2+h_u\\
\Psi^d_{i=n_{Q,1}+1,...,,n^d_{11}}&\overline{\textbf{3}}&\textbf{1}&1/3&-X_Q+(i-1)(x_1-x_2)-h_d\\
\tilde\Psi^d_{i=n_{Q,1}+1,...,,n^d_{11}}&\textbf{3}&\textbf{1}&-1/3&X_Q-(i-1)(x_1-x_2)+x_2+h_d\\
\hline
\multicolumn{5}{|c|}{\text{For $E$'s}}\\
\hline
\Psi^L_{i=1,...,n_{L,1}\leq ,n^e_{11}}&\textbf{1}&\textbf{2}&1/2&-X_L+i(x_1-x_2)+x_2\\
\tilde\Psi^L_{i=1,...,n_{L,1}\leq ,n^e_{11}}&\textbf{1}&\textbf{2}&-1/2&X_L- i(x_1-x_2)\\
\Psi^e_{i=n_{L,1}+1,...,,n^e_{11}}&\textbf{1}&\textbf{1}&1&-X_L+(i-1)(x_1-x_2)-h_d\\
\tilde\Psi^e_{i=n_{L,1}+1,...,,n^e_{11}}&\textbf{1}&\textbf{1}&-1&X_L-(i-1)(x_1-x_2)+ x_2+h_d\\
\hline
\end{array}
\ees
\caption{Gauge charges of the heavy FN fermions}
\label{Heavycharges}
\end{table}
We define $n_{Q/L,1}$ to be the numbers of $SU(2)_W$ doublets pairs in the heavy sector associated to the quark and lepton mass matrices respectively, i.e. which mix with the quark or lepton $SU(2)_W$ doublets of the MSSM (see Figure \ref{treeDiags} to understand how those contribute to the FN mechanism). The subscript $1$ anticipates that there will be equivalent numbers of doublets for each generation. Analogously, there could be heavy pairs of $SU(2)_W$ singlets mixing with the up-type and the down-type quarks or with the electron-like fields, and we denote their numbers by $n_{U,i}$, $n_{D,i}$ and $n_{E,i}$ respectively. In this notation, the total number of heavy pairs needed for generating the Yukawa coupling $h_{ii}^u$ (for example) is
$n_{Q,i} + n_{U,i}$. Said differently, the number of heavy pairs of doublets and singlets are related by relations of the type
\begin{equation}
n_{U,i} = (n_{ii}^u-n_{Q,i}) \theta (n_{ii}^u-n_{Q,i}) \ , \   n_{D,i} = (n_{ii}^d-n_{Q,i}) \theta (n_{ii}^d-n_{Q,i}) \ ,
\label{defNsNonDoublet}
\end{equation}
as clearly depicted on Figure 1.

Those fields together with the MSSM fields form a renormalizable UV theory, with a superpotential formed of (here only for the first generation)
\be
\bead
W\supset&\ \phi_1\tilde\Psi^X_{i}\Psi^X_{i+1} \ (\text{ meaning, e.g., } \phi_1\tilde\Psi^Q_{i}\Psi^Q_{i+1\leq n_{Q,1}}) , \quad \phi_2\tilde\Psi^X_{i}\Psi^X_{i}\ , \\
& \ H_u\tilde\Psi^Q_{n_{Q,1}}\Psi^u_{n_{Q,1}+1}\ , \quad H_d\tilde\Psi^Q_{n_{Q,1}}\Psi^d_{n_{Q,1}+1}\ , \quad H_d\tilde\Psi^L_{n_{L,1}}\Psi^e_{n_{L,1}+1} \ ,
\eead
\label{goodcouplings}
\ee 
where the $\Psi^X$ and $\tilde\Psi^X$ can also be MSSM fields according to the following replacement rules:
\be
Q_{1} \leftrightarrow \tilde\Psi^Q_{0} \ , \quad U_{1} \leftrightarrow \Psi^u_{n^u_{11}+1} \ , \quad D_{1} \leftrightarrow \Psi^d_{n^d_{11}+1} \ , \quad L_{1} \leftrightarrow \tilde\Psi^L_{0} \ , \quad E_{1} \leftrightarrow \Psi^e_{n^e_{11}+1} \ .
\label{replace}
\ee
Those couplings are (generically) the only ones one can write at renormalizable order (see appendix \ref{fermionCouplingsAppendix}) and they are precisely the ones needed to generate \eqref{chiralFN}, via diagrams such as the one of Figure \ref{treeDiags}. 

Mixings to other generations can be similarly implemented via couplings between e.g. $Q_{i>1}$ and one of the $(\phi_1)\Psi^Q$ (again, see Figure \ref{treeDiags} for an example of a diagram which results). However, in order to have mass matrices of rank 3 each, we need to supplement the FN fields of Table \ref{Heavycharges} by their equivalent for the second and third families (see e.g. \cite{Leurer:1992wg,Calibbi:2012yj}), in which case the indices $i$ in Table \ref{Heavycharges} range between $1$ and $n^u_{22},n^d_{22},n^e_{22}$ for the second family, and between $1$ and $n^u_{33}=0,n^d_{33},n^e_{33}$ for the third one. The charges $X_Q$ and $X_L$ in Table \ref{Heavycharges} should also be replaced by $X_Q-(x_1-x_2)(n^u_{11}-n^u_{21})$ and $X_L-(x_1-x_2)(n^e_{11}-n^e_{21})$ for the second family, or by $X_Q-(x_1-x_2)(n^u_{11}-n^u_{31})$ and $X_L-(x_1-x_2)(n^e_{11}-n^e_{31})$ for the third one.

The contribution of the FN fields to the mixed anomaly coefficients are as follows:
\be
\bead
A_{3,\text{FN}} =& \ x_2 \sum_i (2 n_{Q,i} + n_{U,i} +n_{D,i}) =  \ x_2\Big(2(n_{Q,1}+n_{Q,2}+n_{Q,3})+(n^u_{11}-n_{Q,1}) \theta(n^u_{11}-n_{Q,1})\\
& \qquad\qquad\qquad\qquad\qquad\qquad\quad+(n^d_{11}-n_{Q,1}) \theta(n^d_{11}-n_{Q,1})+ (n^u_{22}-n_{Q,2}) \theta(n^u_{22}-n_{Q,2})\\
& \qquad\qquad\qquad\qquad\qquad\qquad\quad+ (n^d_{22}-n_{Q,2}) \theta(n^d_{22}-n_{Q,2})+(n^d_{33}-n_{Q,3})  \theta(n^d_{33}-n_{Q,3})\Big)\\
A_{2,\text{FN}}=& \ x_2\Big(3(n_{Q,1}+n_{Q,2}+n_{Q,3})+n_{L,1}+n_{L,2}+n_{L,3}\Big)\\
A_{1,\text{FN}}=& \ x_2\bigg(\frac{1}{3}(n_{Q,1}+n_{Q,2}+n_{Q,3})+n_{L,1}+n_{L,2}+n_{L,3} \\
& \quad \ \ +\frac{8}{3}[  (n^u_{11}-n_{Q,1}) \theta(n^u_{11}-n_{Q,1})+ (n^u_{22}-n_{Q,2}) \theta(n^u_{22}-n_{Q,2})] \\
& \quad \ \ +\frac{2}{3}[ (n^d_{11}-n_{Q,1}) \theta(n^d_{11}-n_{Q,1})+ (n^d_{22}-n_{Q,2}) \theta(n^d_{22}-n_{Q,2})+ (n^d_{33}-n_{Q,3}) \theta(n^d_{33}-n_{Q,3})]\\
& \quad \ \ + 2[(n^e_{11}-n_{L,1}) \theta(n^e_{11}-n_{L,1})+ (n^e_{22}-n_{L,2}) \theta(n^e_{22}-n_{L,2})+(n^e_{33}-n_{L,3})\theta(n^e_{33}-n_{L,3})]\bigg) \ ,\\
\eead
\ee
where $\theta(x)$ is the Heaviside step function. Hence, we understand that the integrating out of those FN fields generate in addition to \eqref{chiralFN} the following anomalous axionic term in the superpotential\footnote{For an explicit derivation, see e.g. \cite{Anastasopoulos:2006cz} or the appendix D of \cite{Bonnefoy:2018ibr}.} 
\be
W\supset \int d^2\theta\left(-\frac{A_{3,\text{FN}}}{32\pi^2x_2}\log\left(\phi_2\right)(W^a)^2+...\right) \ ,
\label{FNanomalousW}
\ee
where we only displayed the consequence of the QCD anomaly. This would not happen for a vector-like FN sector. Note that only $\phi_2$ appears in \eqref{FNanomalousW} since it is the field which gives its mass to the heavy sector in our construction.

\subsection{Anomaly-free models}\label{chiralExamples}

The presence of \eqref{FNanomalousW} allows one to build "minimal" models where the fermions which participate in the FN mechanism, meaning those which are necessary to generate the hierarchies in masses and mixings, are sufficient to make the model anomaly-free, providing what could be called a minimal anomaly-free gauged FN model. We will not study thoroughly all possible models which achieve this, but, as proofs of principle, we restrict to two specific models.

The first one, which we call Model A in what follows, has only one singlet field $\phi_2$ (and corresponds to a case where $x_1=0$, hence $\phi_1=M$). It reproduces the following Yukawa matrices
\be
Y^u=\left(\begin{array}{ccc} 
\epsilon^{8}&\epsilon^5&\epsilon^3\\
\epsilon^{7}&\epsilon^4&\epsilon^2\\
\epsilon^{5}&\epsilon^2&1\\
\end{array}\right) , \quad  
Y^d=\left(\begin{array}{ccc} 
\epsilon^4&\epsilon^3&\epsilon^3\\
\epsilon^3&\epsilon^2&\epsilon^2\\
\epsilon&1&1\\
\end{array}\right) , \quad  
Y^e=\left(\begin{array}{ccc} 
\epsilon^4&\epsilon^3&\epsilon^3\\
\epsilon^3&\epsilon^2&\epsilon^2\\
\epsilon&1&1\\
\end{array}\right) \ ,
\label{YukawasChosen}
\ee
which fit well the phenomenological values for quarks and mixings when $\tan\beta$ is large. When the FN superfields do not feature any doublet (i.e. $n_{Q,i}=n_{L,i}=0$, leading to a number of heavy fields derived from Table \ref{Heavycharges} and reminded in Table \ref{fermionsTwoModels}),
\begin{table}[t]
 \begin{adjustbox}{max width=\textwidth}
$\begin{array}{|c|c|c|c|c|c|c|c|c|c|c|c|c|c|c|c|}
\hline
&n_{Q,1}&n_{Q,2}&n_{Q,3}&n_{U,1}&n_{U,2}&n_{U,3}&n_{D,1}&n_{D,2}&n_{D,3}&n_{L,1}&n_{L,2}&n_{L,3}&n_{E,1}&n_{E,2}&n_{E,3}\\
\hline
\text{Model A}&0&0&0&8&4&0&4&2&0&0&0&0&4&2&0\\
\hline
\text{Model B}&4&2&0&4&2&0&0&0&0&0&0&0&4&2&0\\
\hline
\end{array}$
\end{adjustbox}
\caption{Number of heavy fields in the two models discussed in section \ref{chiralExamples}\\The numbers of $SU(2)$-singlet and -doublet heavy fields verify relations such as \eqref{defNsNonDoublet}}
\label{fermionsTwoModels}
\end{table}
choosing $h_u=h_d=0$ and $x_2=-\frac{3(3X_Q+X_L)}{16}$ makes all anomalies vanish (and the $\mu$-term $\mu H_uH_d$ is allowed in the superpotential). This amounts to the usual FN model, with the exception that $\frac{\phi}{M}$ is replaced by $\frac{M}{\phi}$. This model is discussed in \cite{Alonso:2018bcg,Smolkovic:2019jow}. It is interesting to note that this anomaly-free and supersymmetric model has the same field content as the ones which were doomed to be saved by a GS mechanism \cite{Ibanez:1994ig,Jain:1994hd,Binetruy:1994ru,Dudas:1995yu}. If one insists on using $\phi_1$ as a dynamical scalar, it is a pure singlet and there will be terms such as $\phi_1^n$ in the superpotential. There is no light degree of freedom in the FN sector in this scenario, which can be constrained by the running of gauge couplings, as discussed in section \ref{gaugeRunningSection}.

On our second model, called Model B, we impose the condition that the heavy FN fields should respect the qualitatively satisfying gauge coupling unification obtained in the MSSM, which can be obtained if the FN fields contribute to the running of the MSSM gauge couplings as $SU(5)$ multiplets (albeit with different $U(1)_\text{FN}$ charges within a same "$SU(5)$ multiplet"). We thus demand that\footnote{This condition can be rewritten in terms of the standard model anomalies, see appendix \ref{anomaliesUnificationAppendix}.}
\be
\bead
n_{Q,1}+n_{Q,2}+n_{Q,3}=& \ n_{U,1}+n_{U,2}= (n^u_{11}-n_{Q,1}) \theta(n^u_{11}-n_{Q,1})+(n^u_{22}-n_{Q,2}) \theta(n^u_{22}-n_{Q,2})\\
 =& \ n_{E,1}+n_{E,2}+n_{E,3}=\sum_i (n^e_{ii}-n_{L,i}) \theta(n^e_{ii}-n_{L,i}) \ ,\\
n_{L,1}+n_{L,2}+n_{L,3}=& \ n_{D,1}+n_{D,2}+n_{D,3} =\sum_i (n^d_{ii}-n_{Q,i}) \theta(n^d_{ii}-n_{Q,i}) \ .
\eead
\label{unifConditions}
\ee
One can check that we need this time two singlets $\phi_1$ and $\phi_2$, if we insist on not using additional spectator fields beyond the ones which enter the FN mechanism. Choosing $x_1=1,x_2=10$, $h_u=h_d=\frac{9}{2},X_Q=-\frac{67}{2},X_L=-\frac{39}{2}$ and the number of heavy fields again displayed in Table \ref{fermionsTwoModels}, all the anomalies vanish and we obtain the following mass matrices (which reproduce the correct masses and mixings up to two $\cO(\lambda)$ deviations \cite{Dudas:1995yu})
\be
Y^u=\left(\begin{array}{ccc} 
\epsilon^{8}&\epsilon^5&\epsilon^4\\
\epsilon^{7}&\epsilon^4&\epsilon^3\\
\epsilon^{4}&\epsilon&1\\
\end{array}\right) , \quad  
Y^d=\left(\begin{array}{ccc} 
\epsilon^3&\epsilon^3&\epsilon^4\\
\epsilon^2&\epsilon^2&\epsilon^3\\
(\overline\epsilon)&(\overline\epsilon)&1\\
\end{array}\right) , \quad  
Y^e=\left(\begin{array}{ccc} 
\epsilon^4&\epsilon^3&\epsilon^3\\
\epsilon^3&\epsilon^2&\epsilon^2\\
\epsilon&1&1\\
\end{array}\right) \ ,
\label{massMatDeviations}
\ee
where by the the parenthesis in the last row of $Y^d$, we  mean that those entries are forbidden by holomorphy. However, they might be generated after field redefinitions to take care of corrections to the K\"ahler potential \cite{Dudas:1995yu}. We nevertheless leave them in \eqref{massMatDeviations}, since they indicate what we choose for the charges of the different fields. Furthermore, notice that, in order to generate the $(1,3)$ and $(2,3)$ entries of $Y^d$, the heavy sector in Table \ref{Heavycharges} should be modified such that, for instance, the index $i$ for the $d$-quark-like heavy fields of the first generation in Table \ref{Heavycharges} is bounded by $n^{d}_{13}$ instead of $n^{d}_{11}$. 

In this model, the $\mu$-term is forbidden and should be generated from the K\"ahler potential via the Giudice-Masiero mechanism \cite{Giudice:1988yz}, by writing\footnote{We have in this case $\mu\sim m_{3/2}\frac{\langle\phi_{1,2}\rangle^2}{\Lambda^2}\sim\frac{F_X\langle\phi_{1,2}\rangle^2}{M_P\Lambda^2}$, where $F_X$ is a non-vanishing auxiliary field from the SUSY breaking sector. Consequently, in a gravity-mediated SUSY breaking scenario with $F_X\sim 10^{11}$ GeV, we understand that we need $\Lambda$ close to $\langle\phi_{1,2}\rangle$ for a $\mu$-term at the TeV scale.}
\be
K\supset \frac{1}{\Lambda^2}H_uH_d\phi_2\overline{\phi_1} \ .
\label{GMasiero}
\ee

An interesting aspect of this model is that it has a light mode, since out of the two phases of $\phi_1$ and $\phi_2$ only one is absorbed by the $U(1)_\text{FN}$ gauge boson and the last one is left as a physical Nambu-Goldstone boson (GB). This feature is generic of the models with two singlets, so we generally comment on it in section \ref{accidentalAxion}.

\subsection{Constraints from the running of gauge couplings}\label{gaugeRunningSection}

The presence of the heavy FN sector adds to the theory many new particles charged under the SM gauge group, so that the running of the MSSM gauge couplings is strongly modified above their mass. In particular, demanding that the model remains perturbative up to some fundamental scale sets strong constraints on the possible masses for the heavy modes (see e.g. \cite{Calibbi:2012yj}). For concreteness, we look at the specific cases of the two models discussed in section \ref{chiralExamples}.

\begin{figure}
\centering
\includegraphics[scale=0.43]{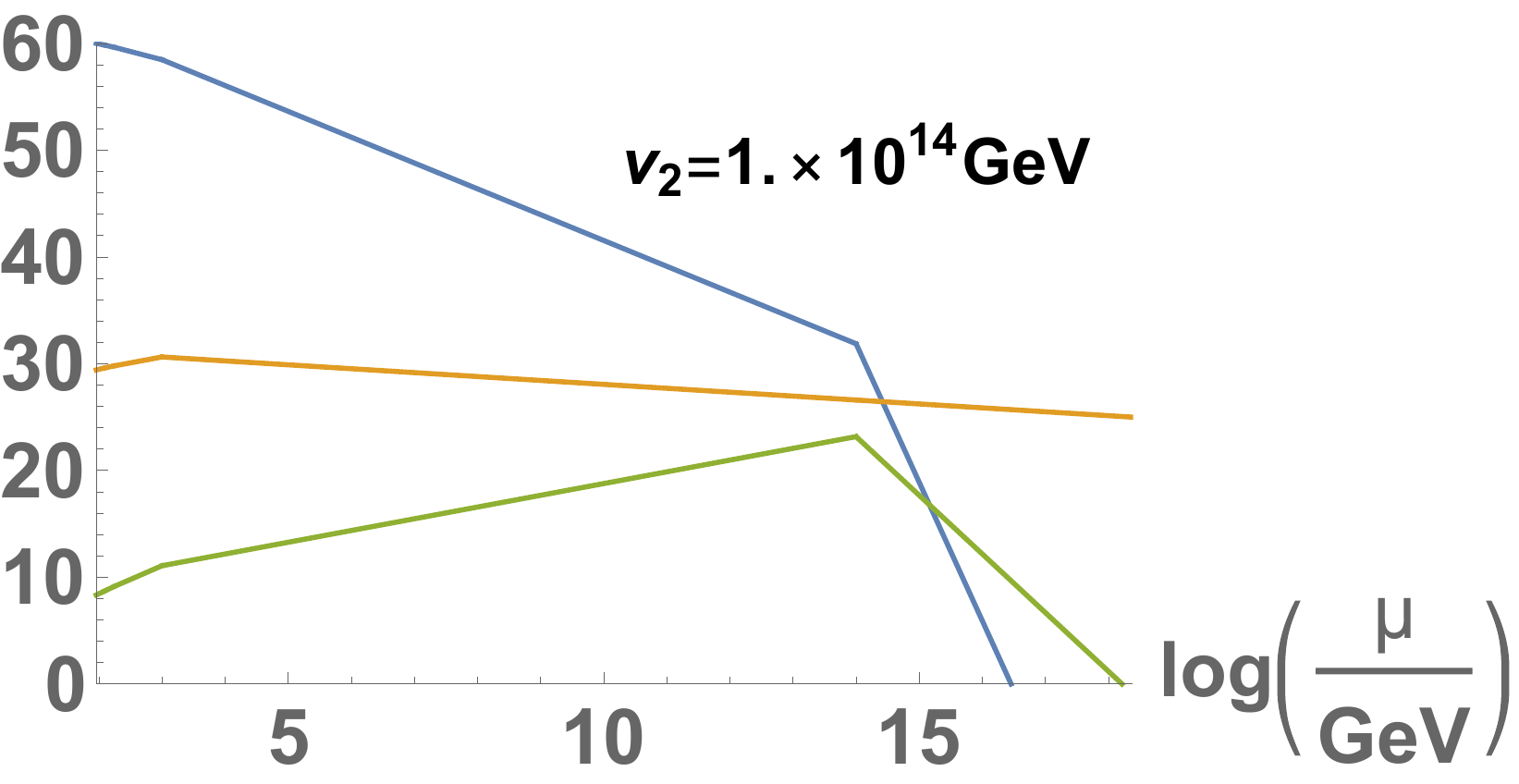}\includegraphics[scale=0.43]{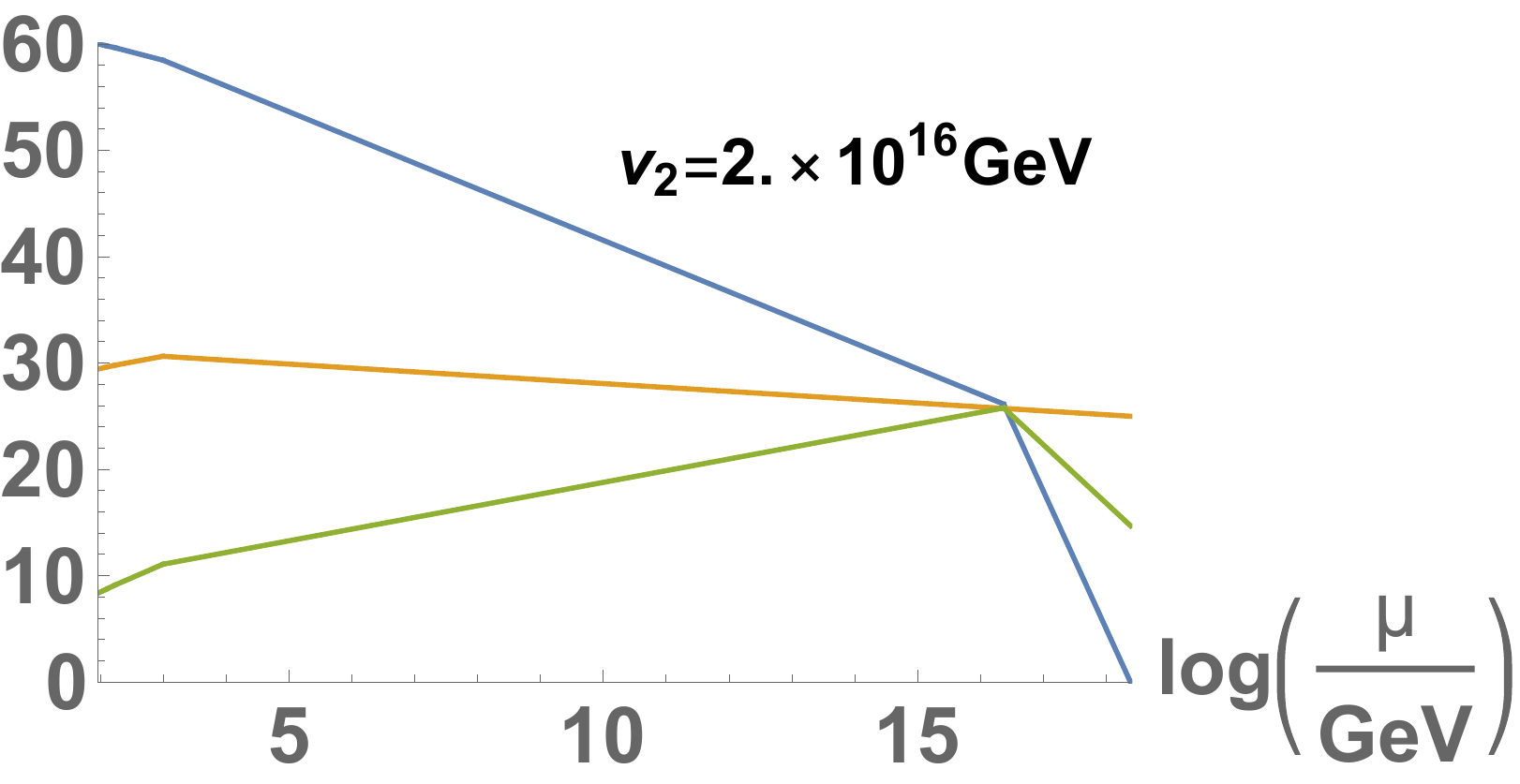}\includegraphics[scale=0.55]{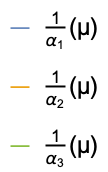}
\caption{Running coupling constants of the MSSM in model A, assuming $m_\text{soft}=$ TeV}
\label{perturbativityRunning}
\end{figure}

Assuming that all the superpartners kick in at a TeV and all the heavy superfields at a high scale $v_2\equiv\langle\phi_2\rangle$, Figure \ref{perturbativityRunning} shows the MSSM gauge couplings running at 1-loop in the model A of section \ref{chiralExamples}, for $v_2=10^{14}$ and $10^{16}$ GeV respectively. We see there that the hypercharge Landau pole, if it is to be above the Planck mass, imposes $v_2\geq10^{16}$ GeV.

\begin{figure}
\centering
\includegraphics[scale=0.43]{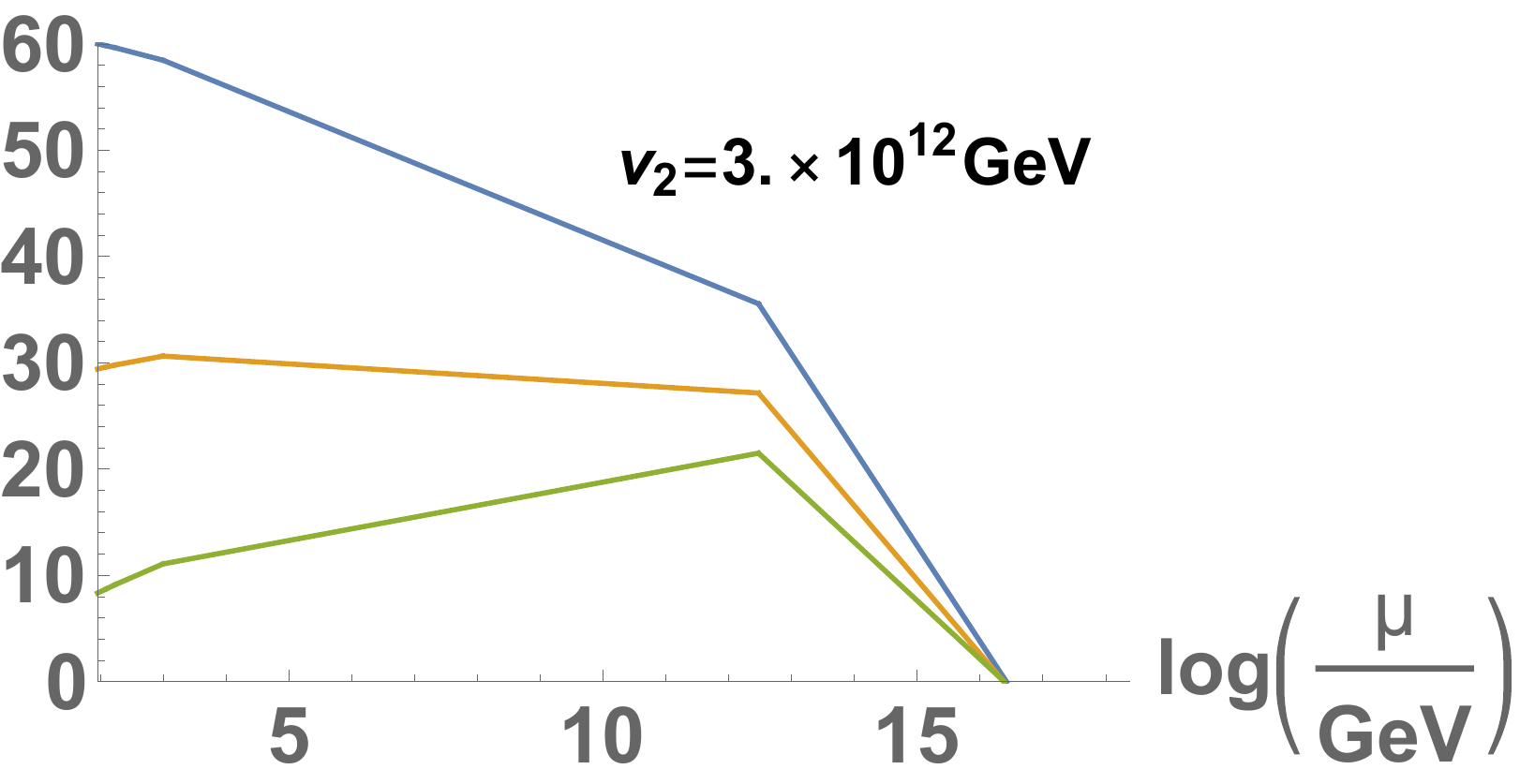}\includegraphics[scale=0.43]{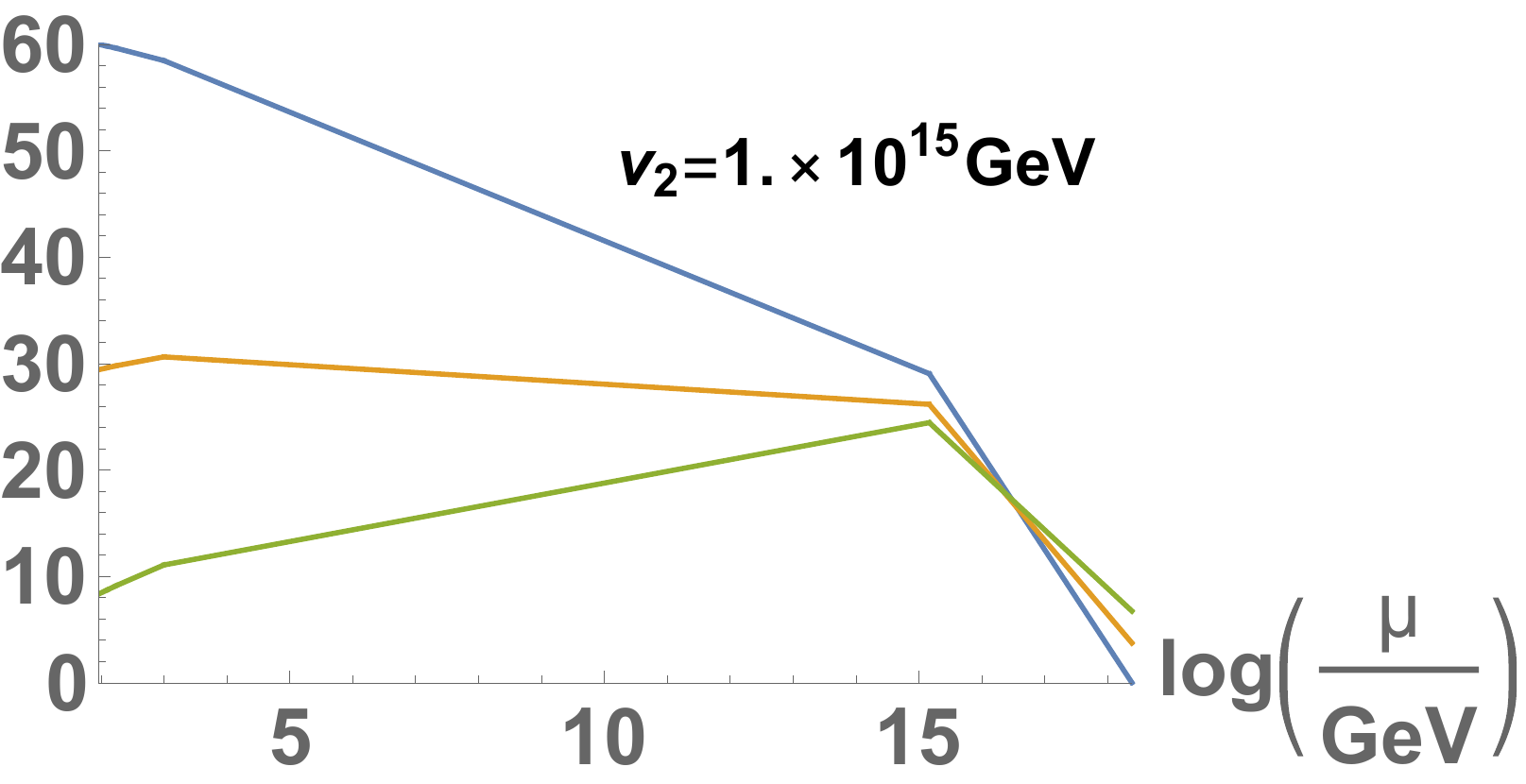}\includegraphics[scale=0.55]{legendRunning.png}
\caption{Running coupling constants of the MSSM in model B, assuming $m_\text{soft}=$ TeV}
\label{perturbativityRunningP}
\end{figure}

With the same assumptions about the supersymmetric spectrum and at 1-loop, Figure \ref{perturbativityRunningP} shows the MSSM gauge couplings running in model B, for $v_2=3\times 10^{12}$ and $10^{15}$ GeV respectively. Here, we see that the hypercharge Landau pole being above the Planck mass imposes $v_2\geq 10^{15}$ GeV. If we instead only impose that the unification happens before any Landau pole, we find that $v_2\geq 3\times10^{12}$ GeV.

Discussing the running of gauge couplings is a good opportunity to emphasize one interesting aspect of chiral models: since their FN sector takes care of both the flavour hierarchies and the anomalies, their heavy field content is expected to be somehow minimal. Consequently, they should be least constrained by the running of gauge couplings. Indeed, the bounds coming from the latter running become stronger when additional charged particles are added to the model, which is necessary if anomalies remain after one integrates the heavy sector which generates the Yukawa couplings. This holds of course for vector-like FN models. 

The minimality of chiral models can be understood as follows: given a mass matrix, one can read off how many heavy fields will be necessary in the FN sector for each generation (for instance, one will need at least $n^u_{11}$ coloured particles to generate the entry $Y^u_{11}$ in a renormalizable model). Quark-like heavy doublets do not overload the model since they are used in both the $U$-like heavy sector and the $D$-like one (unless they are too many such that some of them are only used to generate one Yukawa entry, which happens when $n_{Q,i}>\min(n^u_{ii},n^d_{ii})$). Lepton-like doublets are not minimal in this respect, therefore we can already conclude that if all $n_{L,i}=0$ (and $n_{Q,i}\leq\min(n^u_{ii},n^d_{ii})$), the chiral models we discuss here realize the minimal number of necessary heavy fields. This is the case of models A and B of section \ref{chiralExamples}. Any model using spectator fields to cancel anomalies will have more (or at least as many) heavy SM-charged particles and will be more constrained (maybe marginally) by the running of gauge couplings. If such a model has a physical axion of the kind we discuss later in section \ref{accidentalAxion}, this axion will be less coupled to matter, thus less detectable, than axions originating from chiral models.

For one-singlet models, one can give a clear estimate of how many additional particles would be needed. For instance, in a vector-like counterpart to model A, meaning a model which has one SM singlet and the matrices \eqref{YukawasChosen}, the anomalies from the MSSM(+FN) sector as well as the holomorphy of the supersymmetric couplings impose that one needs at least six pairs of $SU(3)_C$-triplet spectators which contribute to the running of the colour gauge coupling (details can be found in appendix \ref{vectorlikeAppendix}). Such additional particles already have a significant impact on the bounds implied by the $SU(3)_C$ gauge coupling (as illustrated in Figure \ref{perturbativityRunningVectorLike} in appendix \ref{vectorlikeAppendix}). However, in this model they can be singlets under $SU(2)_W$ and without any hypercharge, such that the hypercharge running is unchanged with respect to the chiral case, while it gave the strongest constraint in Figure \ref{perturbativityRunning}. Thus, the chiral model is as much constrained as (or only marginally less constrained than) its vector-like counterpart, although it contains less heavy particles. 

A net strengthening of the bounds arises if we instead try to find a vector-like counterpart to model B. We again leave details to appendix \ref{vectorlikeAppendix}, and we only report here the following result: vector-like counterparts to model B (i.e. vector-like models with spectator fields which preserve the unification of the MSSM gauge couplings) are more constrained than model B itself. For instance, demanding that unification happens within the perturbative regime imposes on those vector-like models that $v_2\geq 4.5\times 10^{13}$ GeV at least, meaning an increase of more than an order of magnitude with respect to model B.

Of course, making such considerations general depend a lot on the $U(1)_\text{FN}$ charges of the Yukawa couplings, as well as on the field content of the theory. For instance, one can find in \cite{Dudas:1995yu} a (two-singlets) model such that the $U(1)_\text{FN}$ charges of the Yukawa couplings are anomaly-free. Hence, a vector-like heavy sector generating them is enough and chiral models do not perform better with respect to the running of gauge couplings. On the other hand, two-singlets chiral models such as the ones we presented come with only two input scales, the vevs of the two singlet scalars, whereas the model aforementioned comes with three: the vevs of the two scalars and the mass scale of the heavy sector, all constrained to reproduce the correct mass hierarchies. In this respect, chiral models have the advantage of minimality.

\subsection{An accidental flavourful Peccei-Quinn symmetry}\label{accidentalAxion}

We now turn to the systematic discussion of the physical GB which arises in models with two singlets $\phi_1$ and $\phi_2$. We stick to the kind of models discussed in sections \ref{chiralGeneral} and \ref{chiralExamples}, namely those where the heavy sector (only or mostly fields participating in the FN mechanism) gets its mass via couplings to $\phi_2$.

The axion is found once we identify the unphysical and the heavy physical CP-odd modes. One pseudoscalar gives the longitudinal component of the $Z$ boson as in usual two Higgs doublets models, and the
pseudoscalar $a_\text{FN}$ which gives the longitudinal component of the $U(1)_\text{FN}$ gauge boson is
\be
a_\text{FN}\propto x_1v_1\theta_1+x_2v_2\theta_2-(h_u+h_d)v_H\theta_H \ ,
\ee
where we wrote $\phi_{1,2}=\frac{r_{1,2}+v_{1,2}}{\sqrt{2}}e^{i\frac{\theta_{1,2}}{v_{1,2}}}$ and $\frac{\theta_H}{v_H}=\arg\left(H_uH_d\right)$, after choosing for simplicity $\langle\abs{H_u}^2\rangle=\langle\abs{H_d}^2\rangle= v_H^2$ and the unitary gauge for the electroweak Goldstone bosons. Two physical pseudoscalars are left behind, one of which will generically be heavy since its mass is unsuppressed in the potential. In what follows, we assume that $\theta_H$ gets a large mass, such that the light physical GB, defined by the orthogonal combination to the gauge and heavy GBs, is mostly made out of the phases of $\phi_1$ and $\phi_2$. This is for instance a valid assumption if the "$b_\mu$" soft term $b_\mu H_uH_d$ is present (i.e. gauge-invariant). Then, the physical leftover GB $a$ is given by
\be
a\propto x_2v_2\theta_1-x_1v_1\theta_2 \ .
\ee
More generally, even if the massive eigenstate is not exactly aligned with $\theta_H$, which is the case for instance in model B, this assumption is still valid at leading order thanks to the large values of $\frac{\langle\phi_{1,2}\rangle}{v_H}$ imposed by the running of the gauge couplings\footnote{For instance, if $h_u+h_d=2x_1$, we expect that $\arg\left(\phi_1^2H_uH_d\right)$ obtains an unsuppressed mass, for instance via supergravity, such that
\be
a \propto x_2v_2\theta_1-x_1v_1\theta_2-\frac{v_H}{v_1}(x_2v_2\theta_H+(h_u+h_d)v_H\theta_2) \ .
\ee}.
Depending on the $U(1)_\text{FN}$ charges of the different scalar fields, the first gauge-invariant operator one could write which violates the shift symmetry of $a$ may be of very high dimension, thus rendering this shift symmetry accidentally protected (more on this below).

We now show that the mode $a$ has couplings similar to the one of flaxions/axiflavons \cite{Wilczek:1982rv,Ema:2016ops,Calibbi:2016hwq,Ema:2018abj}, albeit slightly different numerically, meaning that the family symmetry $U(1)_\text{FN}$ imposes that it has anomalous couplings to gauge fields (and in particular to QCD, making it a Peccei-Quinn axion) and direct couplings to SM fermions.

In the kind of models we consider, the couplings to gluons and photons are completely specified by the mass matrices. Indeed, as already mentioned in \eqref{FNanomalousW}, the heavy sector contributes to the (axionic) anomalous couplings as
\be
W\supset \int d^2\theta\left(-\frac{A_{A,\text{heavy}}}{16\pi^2\cC x_2}\log\left(\phi_2\right)\Tr(W_A^2)\right) \ ,
\label{heavyFNAxionContribution}
\ee
where $A$ refers to either $SU(3)_C$ or $U(1)_\text{em}$, $\cC=1,2$ respectively for a $SU(N)$ or an abelian factor of the gauge group, and we used the fact that all mass terms come from couplings to $\phi_2$. This contribution should be such that its gauge variation precisely cancels that of the contribution from the MSSM fields (here only focusing on QCD):
\be
W\supset \int d^2\theta\left(-\frac{1}{16\pi^2}\log\left(\left(\frac{\phi_1}{\phi_2}\right)^{\sum_i(n^u_{ii}+n^d_{ii})}(H_uH_d)^3\right)\Tr(W_{SU(3)_C}^2)+...\right) \ ,
\ee
Since $\sum_i(n^u_{ii}+n^d_{ii})=\frac{A_{3,\text{SM}}+3(h_u+h_d)}{x_1-x_2}$ and anomaly cancellation imposes $A_{3,\text{heavy}}=-A_{3,\text{SM}}$, we end up with a total contribution
\be
W\supset \int d^2\theta\left(-\frac{A_{3,\text{SM}}}{16\pi^2(x_1-x_2)}\log\left(\phi_1\phi_2^{-\frac{x_1}{x_2}}\left[H_uH_d\left(\frac{\phi_1}{\phi_2}\right)^{\frac{h_u+h_d}{x_1-x_2}}\right]^3\right)\Tr(W_{SU(3)_C}^2)+...\right) \ ,
\label{aQCDcoupling}
\ee
which is obviously gauge-invariant ($a_\text{FN}$ exactly disappears from the $\log$), as it should. 

Besides the coupling to gluons, the heavy chiral fields also feed in the axion-photons coupling. A same line of reasoning gives us the latter:
\be
W\supset \int d^2\theta\left(-\frac{A_\text{em,SM}}{32\pi^2(x_1-x_2)}\log\left(\phi_1\phi_2^{-\frac{x_1}{x_2}}\left[H_uH_d\left(\frac{\phi_1}{\phi_2}\right)^{\frac{h_u+h_d}{x_1-x_2}}\right]^3\right)W_{U(1)_\text{em}}^2+...\right) \ ,
\label{aPhotonCoupling}
\ee
where $A_\text{em,SM}=\frac{A_{1,\text{SM}}+A_{2,\text{SM}}}{2}$ is the MSSM electromagnetic anomaly, so that we understand that
\be
\frac{E}{N}=-\frac{A_{1,\text{SM}}+A_{2,\text{SM}}}{A_{3,\text{SM}}}=\frac{A_{1,\text{heavy}}+A_{2,\text{heavy}}}{A_{3,\text{heavy}}} \ ,
\ee
with the conventions of \cite{Marsh:2015xka}. For instance, model B has $E/N=8/3$, which is the same as in the DFSZ model \cite{Dine:1981rt,Zhitnitsky:1980tq}. Thus, in this respect, our models' predictions are identical to those of usual flaxions/axiflavons. 

To proceed further, let us assume that the bracket term in the $\log$ in \eqref{aQCDcoupling} or \eqref{aPhotonCoupling} precisely corresponds to the heavy pseudoscalar:
\be
a_\text{mass}\propto\frac{\theta_H}{v_H}+\frac{h_u+h_d}{x_1-x_2}\left(\frac{\theta_1}{v_1}-\frac{\theta_2}{v_2}\right) \ .
\ee
This is for instance the case for model B. Then, \eqref{aQCDcoupling} induces a coupling between $a$ and the gluons, since 
\be
\arg\left(\phi_1\phi_2^{-\frac{x_1}{x_2}}\right)\supset \frac{x_2v_2\theta_1-x_1v_1\theta_2}{x_2v_1v_2}=\frac{\sqrt{x_1^2v_1^2+x_2^2v_2^2}}{x_2v_1v_2}a+\cO\left(\frac{v_H}{v_{1,2}}\right) \ ,
\label{aFromLog}
\ee
where we used the canonical normalization for $a$ so that the axion decay constant\footnote{The domain of $a$ is given by $a=a+2\pi f$. In the model defined around \eqref{unifConditions}, $f\equiv\frac{v_1v_2}{\sqrt{v_1^2 +100v_2^2}}\times\min\{\abs{10m-n},(m,n)\in\mathbb{Z}^2\}=\frac{v_1v_2}{\sqrt{v_1^2 +100v_2^2}}$. Thus, $N_\text{DW}=\frac{A_{3,\text{SM}}}{x_2\abs{x_1-x_2}}=2$ in this model.} can be read off from \eqref{aQCDcoupling} and \eqref{aFromLog} (we work at leading order from now on):
\be
f_a=\frac{x_2v_1v_2\abs{x_1-x_2}}{A_{3,\text{SM}}\sqrt{x_1^2v_1^2+x_2^2v_2^2}} \ .
\ee

Dominant couplings between the axion and the SM fermions arise at tree-level from \eqref{chiralFN}, such that the (schematic) coupling between the axion and the SM fermions is as follows:
\be
\cL\supset h_{ij} \left(\frac{\phi_1}{\phi_2}\right)^{n_{ij}}\overline{\psi_{R,j}}\psi_{L,i}H^{(c)}\supset h_{ij} e^{in_{ij} \left(\frac{\theta_1}{v_1}-\frac{\theta_2}{v_2}\right)}\overline{\psi_{R,j}}\psi_{L,i}H^{(c)}\supset h_{ij} e^{i\frac{a}{f_{ij}}}\overline{\psi_{R,j}}\psi_{L,i}H^{(c)} \ ,
\ee 
where we neglected radial degrees of freedom in the first step, and projected the scalar phase onto the physical axion in the second. We also identified the scale of axion-fermions coupling:
\be
f_{ij}=\frac{v_1v_2\sqrt{x_1^2v_1^2+x_2^2v_2^2}}{n_{ij}(x_1v_1^2+x_2v_2^2)} \ ,
\ee
where we see that the axion couples more strongly to lighter generations, since those have larger charges, i.e. larger $n_{ij}$'s. The ratio between the axion coupling to gauge fields $C_a$ and the coupling to fermions $C_{ij}$ is
\be
\frac{C_a}{C_{ij}}\sim\frac{f_{ij}}{f_a}=\frac{A_{a,\text{SM}}}{n_{ij}x_2\abs{x_1-x_2}}\frac{x_1^2v_1^2+x_2^2v_2^2}{x_1v_1^2+x_2v_2^2}\sim\frac{A_{a,\text{SM}}}{n_{ij}x_2\abs{x_1-x_2}}\frac{x_2^2+x_1^2\epsilon^2}{x_2+x_1\epsilon^2} \ .
\label{CoverCchiral}
\ee
As a comparison, flaxion/axiflavon models \cite{Ema:2016ops,Calibbi:2016hwq} find 
\be
\frac{C_a}{C_{ij}}\sim\frac{A_{a,\text{SM}}}{n_{ij}\abs{x_1-x_2}} 
\label{CoverCglobal}
\ee
(where $x_1-x_2$ should be understood as the $U(1)_\text{FN}$ charge of the flavon field). \eqref{CoverCchiral} features qualitative differences with \eqref{CoverCglobal}, for instance it is not only sensitive to $\abs{x_1-x_2}$, which sets the magnitude of the $U(1)_\text{FN}$ charges of the MSSM fields, but also to the absolute value of e.g. $x_2$, such that it contains non-trivial information about the UV physics. Nonetheless, provided the $x_i$'s take reasonable values, the magnitude of \eqref{CoverCchiral} and \eqref{CoverCglobal} are comparable (they are actually equal at order zero in $\epsilon$) and the phenomenological predictions of either kinds of models are qualitatively robust.

An upper bound can actually be imposed on $\langle\phi_{1,2}\rangle$ by requiring that the shift symmetry of the axion $a$ is of high enough quality \cite{PhysRevD.46.539,Kamionkowski:1992mf,Holman:1992us,Fukuda:2017ylt,Bonnefoy:2018ibr} to actually solve the strong CP problem once quantum gravity corrections \cite{Hawking:1987mz,Giddings:1988cx,Banks:2010zn,Harlow:2018jwu,Harlow:2018tng,Fichet:2019ugl} are taken into account. Indeed, we started with gauge symmetries considerations and did not impose any global symmetry on the model. Consequently, we expect to be able to write some gauge invariant operator which would break the shift symmetry of the physical axion. On the other hand, the presence of the $U(1)_\text{FN}$($\times G_\text{SM})$ gauge symmetry may force such an operator to be of very high dimension such that it has no relevant impact on the axion dynamics. 

For instance, in the model B discussed in section \ref{chiralExamples}, the first gauge-invariant operator one could write (beyond those such as \eqref{GMasiero} which carry the heavy axion and respect the light axion shift symmetry by definition) is
\be
c\overline{\phi_2}\phi_1^{10} \ ,
\label{axionMassOperator}
\ee
with $c$ a coupling constant. In the latter case for instance, to be consistent with the measured value of the $\theta$-angle of QCD, $\theta<10^{-10}$ \cite{Baker:2006ts} , we must ensure that:
\be
\bead
\Bigg[m_{a,\text{QCD}} \sim \frac{m_\pi f_\pi}{f_a}\Bigg] &>10^5 \Bigg[m_{a,\text{explicit}}\sim10\sqrt{\abs{c}}\epsilon^4\left(\frac{v_2}{\sqrt{2}}\right)^{\frac{9}{2}}\Bigg] \\
&\text{or equivalently}\\
v_2\lesssim \Big(10^{-5}& \sqrt{2}^{\frac{11}{2}}\abs{c}^{-\frac{1}{2}}\epsilon^{-5}m_\pi f_\pi\Big)^{\frac{2}{11}}\sim2\abs{cM_P^7}^{-\frac{1}{11}}\times10^{11}\text{ GeV} \ ,
\eead
\label{protectQCD}
\ee
where $M_P$ is the reduced Planck mass. We immediately see that this is in tension with the perturbativity bound of section \ref{gaugeRunningSection}, even though not in strict contradiction since there are lots of undetermined order one numbers (e.g. the precise heavy fermion mass or the coefficient $c$). For instance, in our supersymmetric framework, \eqref{axionMassOperator} would be present in the scalar potential if it is also present in the K\"ahler potential and SUSY is broken. Then
\be
c\sim\frac{m_{3/2}^2}{M_P^9} \ ,
\ee
with $m_{3/2}$ the gravitino mass, such that the upper bound in \eqref{protectQCD} can for instance increase by a factor $\sim5\times10^5$ if $m_{3/2}=10^{-4}$ eV, compatible with the gauge mediation of SUSY breaking. \eqref{axionMassOperator}, or its analog in an other model, could also be generated from interference terms between superpotential terms, in which case similar increases of the upper bound may happen.

There is a second gauge-invariant term to be considered in model B, 
\be
H_uH_d\phi_1^{9} \ .
\label{axionMassBis}
\ee 
Understood together with \eqref{axionMassOperator} as unsuppressed potential terms, \eqref{axionMassBis} will never dominate since the weak scale is much smaller than $v_2$. However, since model B is supersymmetric, \eqref{axionMassBis} will appear in the superpotential whereas \eqref{axionMassOperator} lies in the K\"ahler potential, such that a careful analysis of which term contributes dominantly to the axion mass is needed. Nevertheless, in both cases
there is some room for the axion to be a decent QCD axion and the bounds from the running of the gauge couplings to be satisfied\footnote{\eqref{axionMassBis} leads to a scalar potential $\sim m_{3/2}v_H^2\frac{\phi_1^9}{M_P^8}$ so that there are two cases to consider. First, the contribution of \eqref{axionMassOperator} to the axion mass dominates over the one due to \eqref{axionMassBis} if $m_{3/2} > \epsilon\left(\frac{v_H }{v_2}\right)^2 M_P$. With $v_H\sim 10^2$ GeV and using for instance $v_2=10^{12}$ GeV, this means that $m_{3/2}\gtrsim$ MeV. In this case, \eqref{protectQCD} yields $v_2\lesssim 10^{15}$ GeV, which is consistent with our choice for $v_2$ and compatible with the bounds coming from the runnings. On the other hand, if $m_{3/2} < \epsilon\left(\frac{v_H }{v_2}\right)^2 M_P$, \eqref{axionMassBis} dominates and an analysis similar to \eqref{protectQCD} yields $v_2 \lesssim 6.5 \left(\frac{1 \text{ MeV}}{m_{3/2}}\right)^{1/2}10^{11}$ GeV. Achieving more precision demands the inclusion of many unfixed Lagrangian parameters, such as $\tan \beta$.}.

It is then presumably possible to satisfy both kinds of bounds if $v_2\sim 10^{11-13}$ GeV, also implying that explicit breaking of the Peccei-Quinn symmetry could be observable in future experiments aiming at better measuring the neutron (or proton) EDM \cite{Kirch:2013jsa,Anastassopoulos:2015ura}. Furthermore, this value for $v_2$ implies a value for $f_a$ which is compatible with the fact that the flavourful axion makes up part or all of dark matter \cite{Preskill:1982cy,Abbott:1982af,Dine:1982ah} (see also \cite{Marsh:2015xka} for a review), and which is close to the values probed by precision flavour measurements, as discussed now.

\subsection{Flavourful axion phenomenology}\label{axionPhenoSession}

The low-energy phenomenology of our flavourful axion is similar to the one of the flaxion/axiflavon of \cite{Ema:2016ops,Calibbi:2016hwq,Ema:2018abj}. The axion-induced flavour changing transitions of the type $d_i \to d_j + a$, where $d_i$ are $d$-type quarks, and $e_i \to e_j + a$, where $e_i$ are charged leptons, generate decays with the axion in the final state. Experimental limits on such processes set lower limits on the axion decay constant, for a fixed axion-induced flavour changing vertex. Flavour transitions in the quark sector lead to meson decays, the most constraining ones being $K^+ \to \pi^+ + a$ ,   $B^+ \to \pi^+ + a$. The first decay, for example, is bound experimentally \cite{Adler:2008zza} to be $\text{Br}(K^+ \to \pi^+ + a) < 7.3 \times 10^{-11}$, which  leads
to the limit
\begin{equation}
f_a \geq 2 \times 10^{10} \text{ GeV} \times \frac{26}{N_\text{DW}} \left| \frac{(k_v^d)_{12}}{m_s} \right| \ ,
\label{bound1}
\end{equation}
where $k_v^d$ are vector-like fermionic couplings to the axion\footnote{In the notations of section \ref{accidentalAxion}, we have $k^\psi\sim\frac{f_a}{f_{ij}}$, estimated in \eqref{CoverCchiral}.} $\frac{ia}{\sqrt{2} f_a} ( k_a^{\psi} {\bar \psi}_i \gamma_5 \psi_i + k_v^{\psi} {\bar \psi}_i  \psi_i )$ and $N_\text{DW} = \frac{\sum_i (2 q_{Q_i} + q_{U_i} + q_{D_i})}{x_2\abs{x_1-x_2}}$ is the domain wall number.   

Flavour transitions in the charged lepton sector lead to lepton number per species non-conserving processes , most constraining  one being $\mu \to e + a + \gamma$, constrained experimentally
to be $\text{Br}(\mu \to e + a + \gamma) < 1.1 \times 10^{-9}$, which leads to the bound \cite{Goldman:1987hy}
\begin{equation}
f_a \geq 1 \times 10^{8} \text{ GeV}\times \frac{26}{N_{DW}} \left| \frac{(k_v^l)_{12}}{m_{\mu}} \right| \ .
\label{bound2}
\end{equation}
Our perturbativity bounds due to the running effects of heavy fields $v_2 \gtrsim 10^{12}$ GeV are compatible with all these bounds. However, one-two orders of magnitude improvement of experimental data in the near future is expected and will start probing our models\footnote{As we emphasized in section \ref{gaugeRunningSection}, for a given vector-like gauged FN model with an axion, the bound on $v_2$ may increase (perhaps weakly) with respect to those derived in chiral models, so that the axion is in principle less detectable than ones from chiral models.}. 

Another source of flavour violation is the coupling  of quarks and leptons to the $U(1)_\text{FN}$ gauge boson  $Z'$. Baryon number cannot be violated in this way, otherwise the flavon would carry baryon number. Lepton number per species could be violated, but   due to the high scale of $U(1)_\text{FN}$ symmetry breaking $v_2 \gtrsim 10^{12}$ GeV,  $Z'$-induced lepton number non-conserving processes are currently unobservable. 

\subsection{Non-supersymmetric models}\label{nonSUSY}

We now briefly comment on non-SUSY chiral FN models, by again explicitly displaying such models for definiteness. We focus for simplicity on holomorphic models with two Higgs doublets, meaning that \eqref{chiralFN}, once complemented by its hermitian conjugate, now defines the lagrangian of the theory, with $H_{u,d}$ referring to scalar fields and $Q_i,U_j,D_j,E_j$ to left-handed Weyl fermions. The heavy FN fields are similarly all fermionic, except $\phi_{1,2}$ which are scalars.

\begin{figure}
\centering
\includegraphics[scale=0.43]{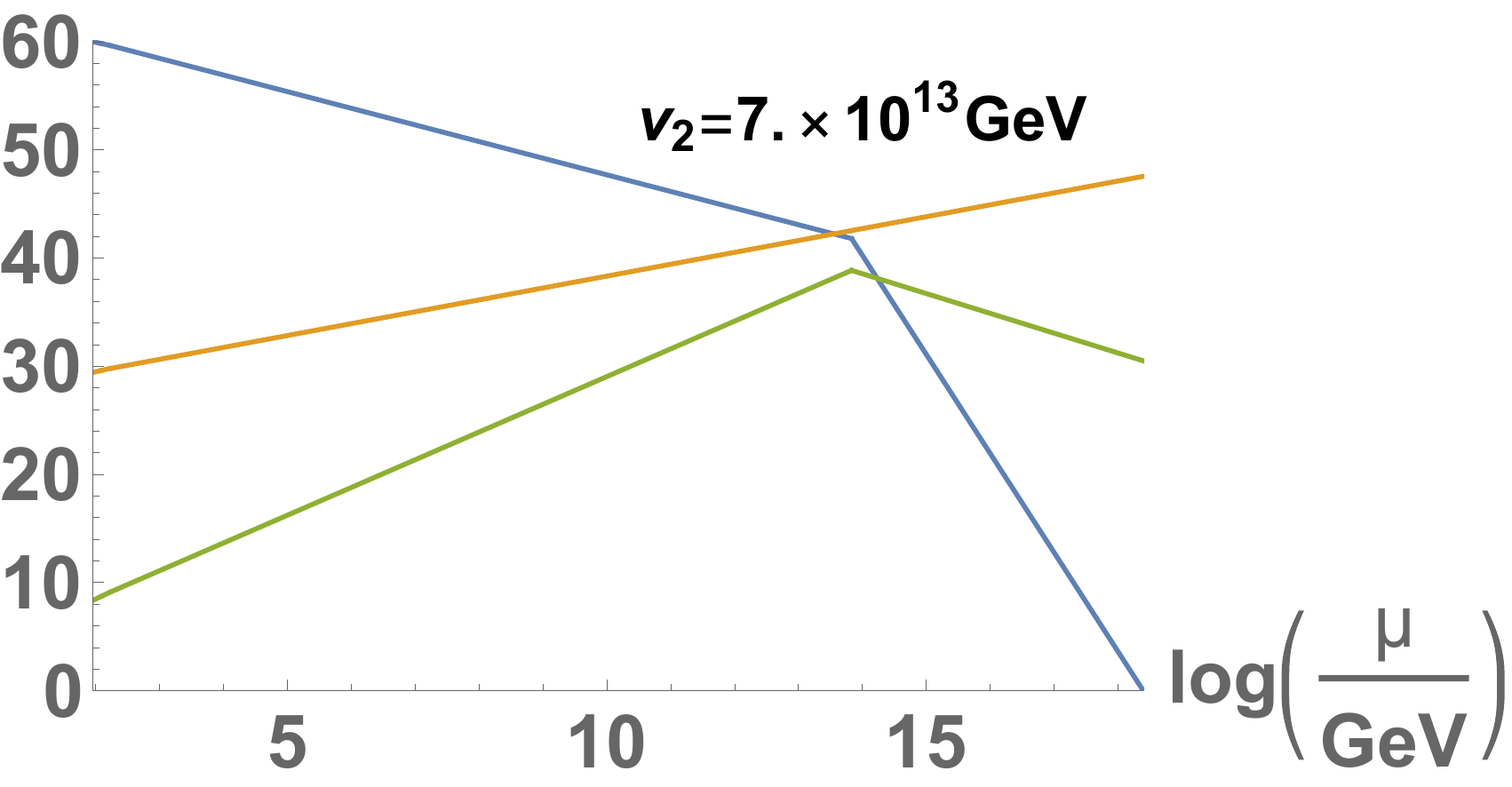}\includegraphics[scale=0.43]{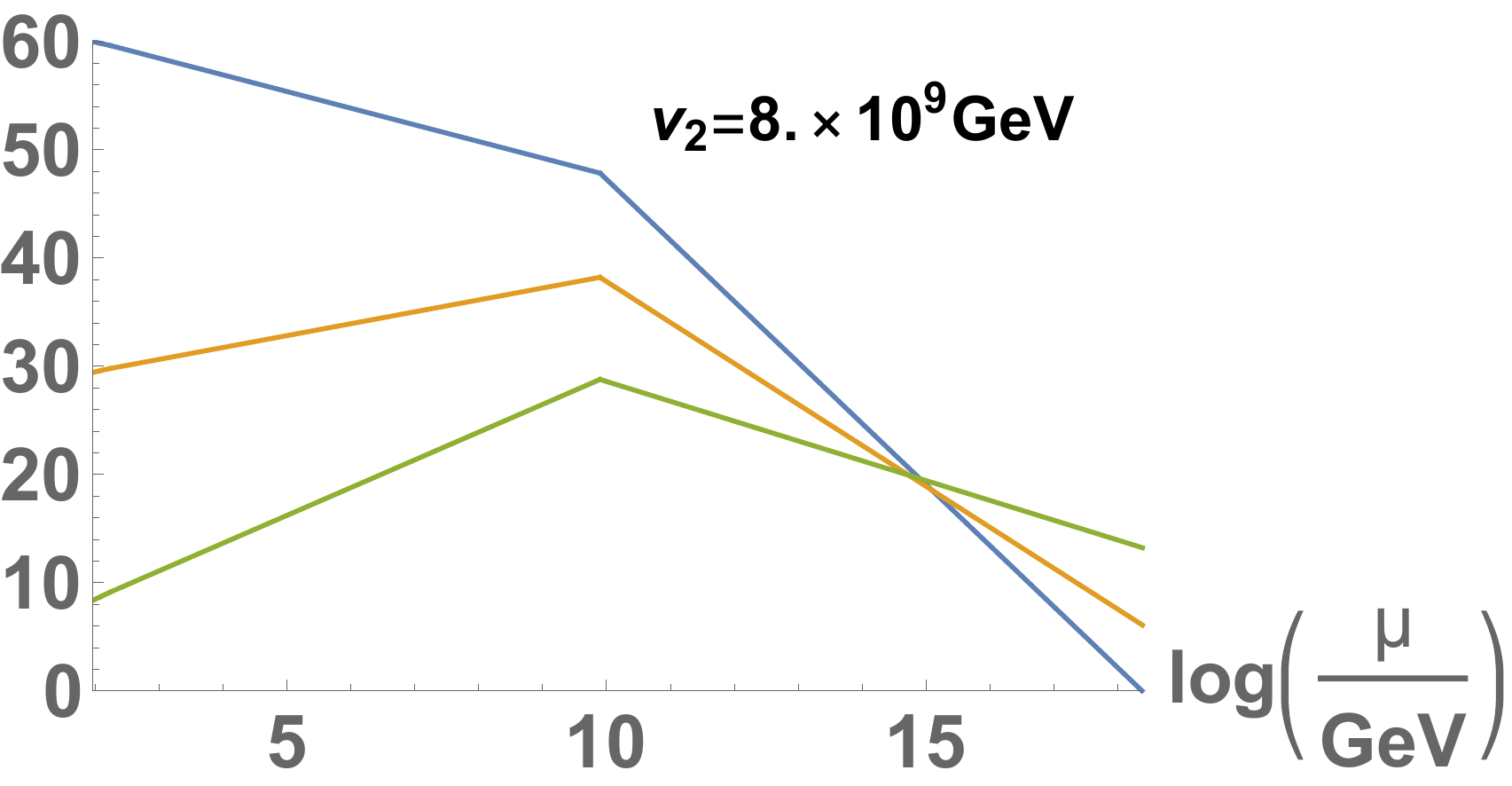}\includegraphics[scale=0.55]{legendRunning.png}
\caption{Running coupling constants of the SM\\Left panel: first model discussed in section \ref{nonSUSY}, right panel: second model}
\label{perturbativityRunningNonSUSY}
\end{figure}

It is straightforward to check that the model A of section \ref{chiralExamples} is also valid as a non-SUSY model, consistently with the findings of \cite{Alonso:2018bcg,Smolkovic:2019jow} (indeed, higgsinos carry no charge under $U(1)_\text{FN}$, so they can be removed at no cost, and all the heavy fermions remain)\footnote{Since the Higgs fields carry no $U(1)_\text{FN}$ charge, one of them can be discarded by defining e.g. $H_u=H_d^c$.}. For the Yukawa matrices in \eqref{massMatDeviations}, there are again models with two charged scalars, for instance if $x_1=1,x_2=\frac{1}{2}$, $h_u=h_d=3,X_Q=0,X_L=-\frac{1}{6}$ and $n_{Q,1}=5, n_{Q,2}=1, n_{Q,3}=0, n_{L,1}=0, n_{L,2}=2, n_{L,3}=0$ (so that $n_{U,1}=3, n_{U,2}=3, n_{U,3}=0,n_{D,1}=0, n_{D,2}=1, n_{D,3}=0,n_{E,1}=4, n_{E,2}=0, n_{E,3}=0$). With such charges the physical pseudoscalar contained in $\phi_{1,2}$, which has a QCD axion-like coupling to gluons, is too heavy to be a proper QCD axion since there can be operators such as  
\be
\phi_1\overline{\phi_2}^2 \ 
\ee
in the lagrangian. On the other hand, the "$\mu$-term" has the following form
\be
H_uH_d\phi_1^2\left(\frac{\phi_1}{\Lambda}\right)^4 \ ,
\ee
with $\Lambda$ a high scale, so that the pseudoscalar in $H_{u,d}$ is heavier than e.g. a TeV if
\be
v_1^2\left(\frac{v_1}{\Lambda}\right)^4>\text{ TeV}^2 \ ,
\ee
i.e.
\be
v_1>\left(\text{TeV}\times\Lambda^2\right)^{\frac{1}{3}}\approx\big\vert_{\text{if }\Lambda=M_P}2\times10^{13} \text{ GeV} \ .
\ee
This may give a stronger bound on $v_{1,2}$ than the running of the SM gauge couplings: the latter is indeed slightly less severe than the ones seen in section \ref{gaugeRunningSection} for SUSY models. Figure \ref{perturbativityRunningNonSUSY} shows the running for the two models aforementioned in this section, and we see there that the heavy sector masses for which the hypercharge gauge coupling blows up at the Planck scale are reduced by a few orders of magnitude with respect to what appears on Figures \ref{perturbativityRunning} and \ref{perturbativityRunningP}.

\section{Conclusion}

We studied the gauging of a horizontal abelian symmetry generating the Froggatt-Nielsen mechanism, when the heavy fields in the UV completion of the mechanism are chiral with respect to this family symmetry. This for instance happens when the small parameter which explains the flavour hierarchies is composed of the vevs of two charged scalar fields which respectively mix and give masses to the heavy sector. The mixed anomalies between the Standard Model gauge group and the new symmetry are modified in this setup, such that the anomaly-free completions of the model are not the same as in the usual case when the heavy sector is vector-like.

We mostly focused on supersymmetric models, since their holomorphy properties usually do not leave much freedom for the anomalies to cancel. Unlike the vector-like heavy sector case, for which it has been shown that the minimal embedding of the FN symmetry is always anomalous at the level of the MSSM, with our chiral heavy sectors the mixed anomalies are enough disentangled from the mass matrices so that they sometimes vanish without adding a Green-Schwarz mechanism or any other spectator field than the ones which are necessary for the FN mechanism to take place. We gave specific examples where this "minimal" UV content is realized, and compared them to gauged vector-like models, with spectator fields cancelling the anomalies, to precisely illustrate what we mean by "minimal": chiral models push Landau poles to the highest possible values, so that bounds on the input scales are the loosest possible, and they minimize the number of input scales in the problem. We also presented non-supersymmetric examples with the same behaviour.

Moreover, we emphasized the fact that chiral models often come with a physical axion mode, which has couplings typical of a QCD flaxion/axiflavon. In such models, the gauging of the FN symmetry makes it easy to protect the axion mass, which is a significant difference with respect to flavourful axions originating from a global FN mechanism. The qualitative axion phenomenology is similar to the one of global flaxion/axiflavon models, meaning that the axion couplings are mainly dictated by low-energy physics, which remains as a robust prediction. However, there are slight changes in the axion couplings to gauge fields, since the latter are already generated by the integrating-out of heavy FN fermions, not only by the SM ones. In addition, irrespective of the model, strong bounds on the input scales describing the heavy sector can be derived from the running of the (MS)SM gauge couplings, which imposes that the scale of spontaneous FN symmetry breaking is at least intermediate ($10^{12-13}$ GeV). Those lower bounds are stricter in gauged vector-like FN models, so that the axions arising in chiral models are maximally coupled/detectable. Nonetheless, such bounds are enough for the models to be automatically compatible with experimental results on flavour-changing processes, albeit not too high so that one will start scanning the couplings of the flavourful axions after the experimental sensitivity increases slightly. Finally, since the SM mass matrices make anomaly cancellation compatible with the MSSM gauge coupling unification, one can define anomaly-free models which preserve gauge coupling unification such that the couplings of the axion, anomaly cancellation and the gauge couplings running are all entangled.

\section*{Acknowledgments}

Q.B. would like to thank the Institute of Theoretical Physics at the University of Warsaw for their financial support and their kind hospitality. E.D. was supported in part by the Agence Nationale de la 
Recherche project ANR Black-dS-String.
 S.P. thanks the Instituto de Fisica Teorica (IFT UAM-CSIC) in Madrid for its support via the Centro de Excelencia Severo Ochoa
Program under Grant SEV-2016-0597  and Belen Gavela and Pablo Quilez  for very useful discussions. The S.P. research was partially supported by the Munich Institute for Astro- and Particle Physics (MIAPP) of the DFG Excellence Cluster Origins (www.origins-cluster.de).

\begin{appendices}

\section{All possible superpotential terms}\label{fermionCouplingsAppendix}

In this appendix, we study the most general renormalizable couplings which can appear in the superpotential for the kind of models discussed in section \ref{sectionMain}. We assume that $x_1-x_2\neq 0$, so that the Froggatt-Nielsen mechanism is operating, as well as R-parity. 

With the matter content of Table \ref{Heavycharges} (together with the replacement rules \eqref{replace}), Table \ref{bilinears} displays the renormalizable invariant-under-$G_\text{SM}$ terms  which one may write down in the superpotential if they are also $U(1)_\text{FN}$-invariant (i.e. if they are invariant under the full gauge group).
\begin{table}
\bes
\begin{array}{|c|c|c|}
\hline
G_\text{SM}\text{ invariant term}&U(1)_\text{FN}\text{ charge}&\text{Invariant if}\\
\hline
&&\\
\left.\begin{matrix}\Psi^Q_{i}\tilde\Psi^Q_{j}\\\Psi^u_{i}\tilde\Psi^u_{j}\\\Psi^d_{i}\tilde\Psi^d_{j}\\\Psi^L_{i}\tilde\Psi^L_{j}\end{matrix}\right \}&(i-j)x_1-(i-j-1)x_2&\begin{cases}i=j\text{ and }x_2=0\\\qquad\qquad\text{or }\\i\neq j\text{ and }x_1=\frac{(i-j-1)x_2}{i-j}\end{cases}\\
&&\\
\left.\begin{matrix}\Psi^Q_{i}\tilde\Psi^Q_{j}\\\Psi^u_{i}\tilde\Psi^u_{j}\\\Psi^d_{i}\tilde\Psi^d_{j}\\\Psi^L_{i}\tilde\Psi^L_{j}\end{matrix}\right \}\times\phi_1&(i-j-1)(x_1-x_2)&i=j+1\\
&&\\
\left.\begin{matrix}\Psi^Q_{i}\tilde\Psi^Q_{j}\\\Psi^u_{i}\tilde\Psi^u_{j}\\\Psi^d_{i}\tilde\Psi^d_{j}\\\Psi^L_{i}\tilde\Psi^L_{j}\end{matrix}\right \}\times\phi_2&(i-j)(x_1-x_2)&i=j\\
&&\\
\left.\begin{matrix}\tilde\Psi^Q_{i}\Psi^u_{j}H_u\\\tilde\Psi^Q_{i}\Psi^d_{j}H_d\\\tilde\Psi^L_{i}\Psi^e_{j}H_d\end{matrix}\right \}&(j-i-1)(x_1-x_2)&i=j-1\\
&&\\
\left.\begin{matrix}\Psi^Q_{i}\tilde\Psi^u_{j}H_d\\\Psi^Q_{i}\tilde\Psi^d_{j}H_u\\\Psi^L_{i}\tilde\Psi^e_{j}H_u\end{matrix}\right \}
&(i-j+1)x_1-(i-j-1)x_2+h_u+h_d&\begin{cases}i=j-1\text{ and }x_2=\frac{h_u+h_d}{i-j-1}\\\qquad\qquad\qquad\text{or }\\i\neq j-1\text{ and }x_1=\frac{(i-j-1)x_2-h_u-h_d}{i-j+1}\end{cases}\\
&&\\
\hline
\end{array}
\ees
\caption{$G_\text{SM}$-invariant superpotential terms formed with superfields defined in section \ref{sectionMain}}
\label{bilinears}
\end{table}
There, the three sets of possibilities in the middle of the Table are those which are required to implement the FN mechanism as emphasized in \eqref{goodcouplings}. They can always be written down provided the values of the subscripts $i$ and $j$ are chosen appropriately. The other two can appear when $x_1,x_2,h_u$ and $h_d$ verify specific arithmetical relations. For instance, they are not allowed in the two models discussed in section \ref{chiralExamples}.

\section{Anomalies and unification}\label{anomaliesUnificationAppendix}

In this appendix, we show that the "unification" relations \eqref{unifConditions} can be reexpressed in terms of the mixed anomalies in the MSSM sector. To see this, let us recall (see the discussion around \eqref{heavyFNAxionContribution} in section \ref{accidentalAxion}) that the contribution to the axion couplings generated by the integrating out of the heavy FN sector is:
\be
W\supset \int d^2\theta\left(-\frac{A_{A,\text{heavy}}}{16\pi^2\cC x_2}\log\left(\phi_2\right)\Tr(W_A^2)\right) \ ,
\ee
where again $\cC=1,2$ respectively for a $SU(N)$ or an abelian factor of the gauge group and, with our conventions, $\frac{A_{A,\text{heavy}}}{x_2}$ counts the heavy chiral fields which are charged under the gauge factor $A$ (with multiplicity and charge squared for an abelian gauge factor). Thanks to the holomorphy in our SUSY model, the same anomaly coefficients appear in the $\beta$-function for the gauge couplings:
\be
\frac{1}{g_A^2(\mu)}=\frac{1}{g_A^2(\mu_0)}+\frac{b_A^\text{MSSM}}{8\pi^2}\log(\frac{\mu}{\mu_0})-\frac{A_{A,\text{heavy}}}{64\pi^2}\log(\frac{\mu}{v_2}) \ .
\ee
Respecting the gauge unification of the MSSM thus demands 
\be
A_{SU(3)_C,\text{heavy}}=A_{SU(2)_W,\text{heavy}}=\frac{3}{5}A_{U(1)_Y,\text{heavy}} \ ,
\ee
which, for anomaly-free models, is extended to
\be
A_{SU(3)_C,\text{SM}}=A_{SU(2)_W,\text{SM}}=\frac{3}{5}A_{U(1)_Y,\text{SM}} \ .
\ee
This relation agrees well with phenomenological mass matrices, and is the one required to implement the Green-Schwarz mechanism \cite{Ibanez:1994ig,Jain:1994hd,Binetruy:1994ru,Dudas:1995yu}, consistently with the fact that the phase $\theta_2$ of $\phi_2$ does generate a GS mechanism here once the heavy fields are integrated out. 

As a consistency check, it is straightforward to check that model B of section \ref{chiralExamples} indeed verifies
\be
\bead
A_{SU(3)_C,\text{SM}}&=A_{SU(2)_W,\text{SM}}=\frac{3}{5}A_{U(1)_Y,\text{SM}}\\
&=-A_{SU(3)_C,\text{heavy}}=-A_{SU(2)_W,\text{heavy}}=-\frac{3}{5}A_{U(1)_Y,\text{heavy}}=-180 \ .
\eead
\ee

\section{Minimal vector-like models}\label{vectorlikeAppendix}

We detail in this appendix the construction of and the bounds on the vector-like counterparts to model A and B.

\begin{figure}
\centering
\includegraphics[scale=0.5]{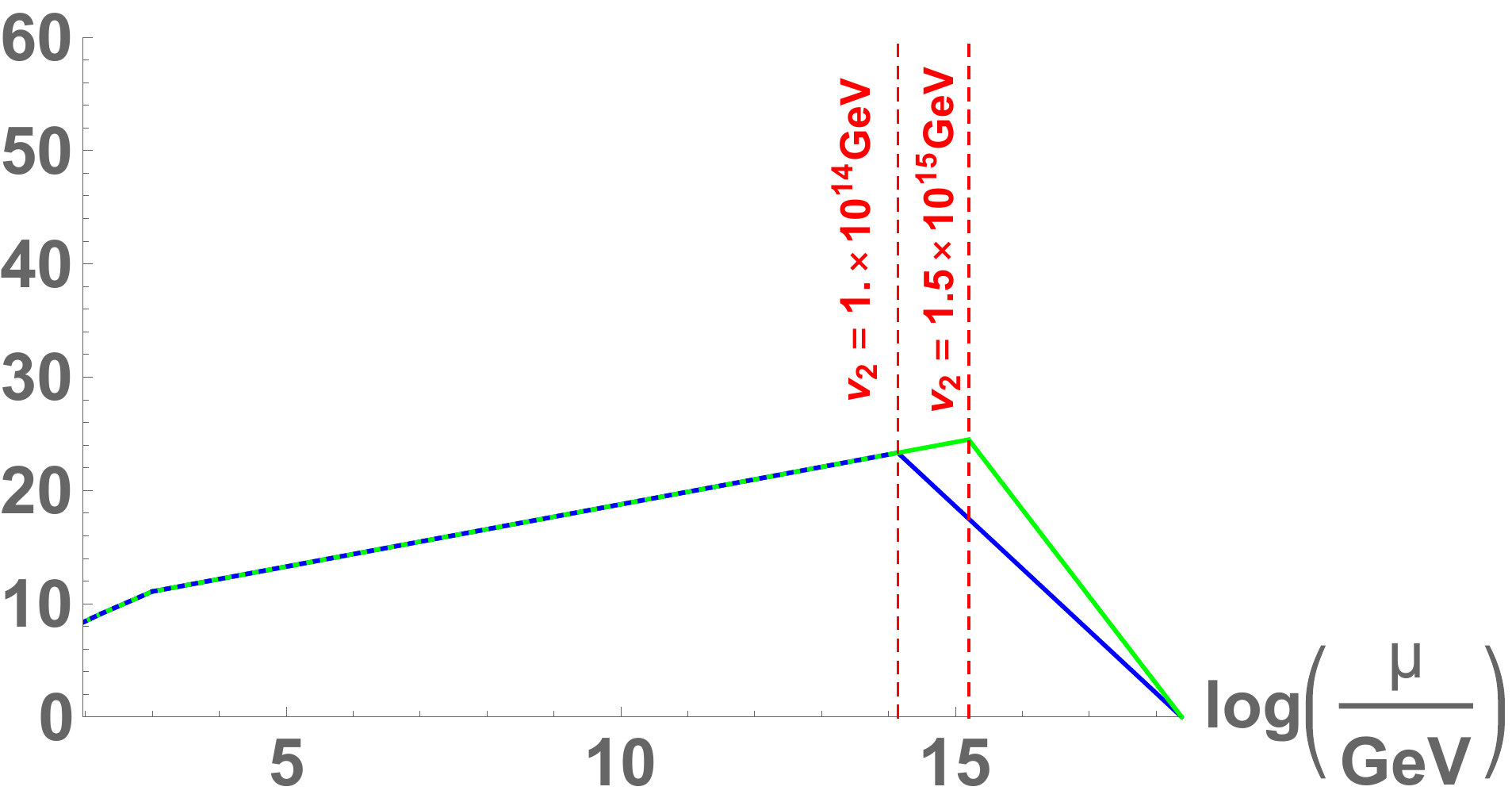}\includegraphics[scale=0.55]{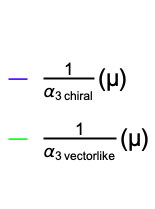}
\caption{Running coupling constant of $SU(3)_C$ in model A when $v_2=10^{14}$ GeV,\\ and in its gauged vector-like counterpart with six triplets when $v_2=1.5\times10^{15}$ GeV, assuming $m_\text{soft}=1$ TeV}
\label{perturbativityRunningVectorLike}
\end{figure}

A vector-like counterpart to model A is a model which has one SM singlet (which amounts to the choice $x_1=1$ and $x_2=0$) and the matrices \eqref{YukawasChosen}. Thus, the anomalies from the MSSM(+FN) sector are
\be
\bead
A_{3,\text{SM}} =& \ 18 - 3 (h_u + h_d)\\
A_{2,\text{SM}}=& \ -16+3(3X_Q+X_L)+h_u + h_d\\
A_{1,\text{SM}}=& \ 64-3(3X_Q+X_L) -7(h_u + h_d)\\
A'_{1,\text{SM}}=& \ -128 +X_Q(6h_d-12h_u+36)+X_L(6h_d-12)+64h_u-40h_d+5(h_d^2-h_u^2) \ ,
\eead
\label{SManomaliesVectorModelA}
\ee
the last coefficient referring to the $U(1)_Y\times U(1)_\text{FN}^2$ anomaly. Restricting ourselves to spectators $(\chi^i,\tilde\chi^i)$ which get their mass from couplings to the singlet $\phi_1$ as follows
\be
W \supset c_{i}\phi_1\chi^i\tilde \chi^i \ ,
\ee
and which live in the singlet or fundamental representations of the SM gauge groups (as shown in Table \ref{spectatorFiedsArray}),
\begin{table}[h]
\centering
$\begin{array}{|c|c|c|c|c|}
\hline
&SU(3)_C&SU(2)_W&U(1)_Y&U(1)_\text{FN}\\
\hline
\chi^i&\textbf{3} \text{ or } \textbf{1}&\textbf{2} \text{ or } \textbf{1}&y_{\chi^i}&q_{\chi^i}\\
\tilde\chi^i&\overline{\textbf{3}} \text{ or } \textbf{1}&\textbf{2} \text{ or } \textbf{1}&-y_{\chi^i}&-q_{\chi^i}+1\\
\hline
\end{array}$
\caption{Gauge charges of the spectator fields}
\label{spectatorFiedsArray}
\end{table}
their contribution to the anomalies are
\be
\bead
A_{3,\text{spect.}} =& \ \sum_i(1+\delta^i_2)\delta^i_3\\
A_{2,\text{spect.}}=& \ \sum_i(1+2\delta^i_3)\delta^i_2\\
A_{1,\text{spect.}}=& \ \frac{1}{2}\sum_i(1+\delta^i_2)(1+2\delta^i_3)y_{\chi^i}^2\\
A'_{1,\text{spect.}}=& \ \frac{1}{2}\sum_i(1+\delta^i_2)(1+2\delta^i_3)y_{\chi^i}[q_{\chi^i}^2-(q_{\chi^i}-1)^2]  \ ,
\eead
\label{SpectatorAnomaliesVectorModelA}
\ee
where $\delta^i_{2/3}$ is equal to one if the corresponding spectator field is in the fundamental representation of $SU(3)_C/SU(2)_W$, and zero if it is a singlet. In particular, $A_{1-2-3,\text{spect.}}$ are positive, and $A_{2-3,\text{spect.}}$ are integer, such that $A_{1-2-3,\text{SM}}$ should be negative and $A_{2-3,\text{SM}}$ should be integer for the anomalies to cancel. One can also write
\be
A_{1,\text{SM}}+A_{2,\text{SM}}=12+2A_{3,\text{SM}} \ ,
\ee 
such that $A_{1,\text{SM}}+A_{2,\text{SM}}\leq0\implies A_{3,\text{SM}}\leq-6$, which in turn implies that at least six pairs of spectator triplets will contribute to the running of the colour gauge coupling. Such additional particles already have a significant impact on the bounds implied by the $SU(3)_C$ gauge coupling, as illustrated in Figure \ref{perturbativityRunningVectorLike}.
All the anomalies can be cancelled here by choosing\footnote{We also see in passing that the $\mu$-term cannot be gauge-invariant in such a one-singlet vector-like model, since $h_u+h_d=0\implies A_{3,\text{SM}}>0$.} $h_u=h_d=4$, $X_Q=0$, $X_L=\frac{8}{3}$ and by adding exactly six pairs of spectator coloured triplets, singlets under $SU(2)_W$ and without any hypercharge. In this case, the hypercharge running is unchanged with respect to the chiral case, while it gave the strongest constraint in Figure \ref{perturbativityRunning}, so that the chiral model is as much constrained as (or only marginally less constrained than) its vector-like counterpart, although it contains less heavy particles.

Let us now turn to the derivation of the bounds on vector-like counterparts to model B, meaning models which have two SM singlets of $U(1)_\text{FN}$ charges $-1$ and $-10$, the matrices \eqref{massMatDeviations} and $SU(5)$-like unification as in \eqref{unifConditions}. Assuming that the charges of the MSSM particles are the same as in model B and that the Yukawa couplings are all expressed in terms of the same singlet, then normalizing all the $U(1)_\text{FN}$ charges such that the Yukawa couplings have natural integer charges (i.e. using \eqref{newSManomalies} with $x_1-x_2=1$), the MSSM anomalies are
\be
\bead
A_{3,\text{SM}} =& \ 17 - 3 (h_u + h_d)\\
A_{2,\text{SM}}=& \ -19+3(3X_Q+X_L)+h_u + h_d\\
A_{1,\text{SM}}=& \ \frac{199}{3}-3(3X_Q+X_L) -7(h_u + h_d)\\
A'_{1,\text{SM}}=& \ -135 +X_Q(6h_d-12h_u+38)+X_L(6h_d-12)+68h_u-40h_d+5(h_d^2-h_u^2)\ .
\eead
\label{SManomaliesVectorModelA}
\ee
With our assumptions and normalizations, model B contains either a SM singlet of $U(1)_\text{FN}$ charge $-1$ and an other one of charge $-10$, or a SM singlet of $U(1)_\text{FN}$ charge $-1$ and an other one of charge $-\frac{1}{10}$. $A_{1-2-3,\text{SM}}$ should be negative in both cases, and $A_{2-3,\text{SM}}$ should be either integer or multiples of $\frac{1}{10}$, for the same reasons as presented previously. Since the running of the hypercharge coupling gives the strongest constraint in Figure \ref{perturbativityRunningP}, the bound it induces will strengthen if at least a spectator field has a non-vanishing hypercharge\footnote{The bound could also increase due to contributions of the other gauge couplings.}. Assuming the opposite would imply $A_{1,\text{spect.}}=A_{1,\text{SM}}=0$, which gives (since $A_{1,\text{SM}}+A_{2,\text{SM}}-2A_{3,\text{SM}}=\frac{40}{3}$)
\be
A_{2,\text{SM}}-2A_{3,\text{SM}}=\frac{40}{3} \ ,
\ee
whereas this quantity should be a multiple of $1$ or $\frac{1}{10}$. We thus conclude that one needs at least one spectator with a non-zero hypercharge, which would in turn make the running of the hypercharge coupling a bit steeper than it was in the chiral model\footnote{This conclusion might be evaded, and the running brought back to the one of the chiral model, by relaxing some of the assumptions we made, for instance if one starts using both singlets to generate the mass matrices, contrary to what we assumed here. On the other hand, this means that the $U(1)_\text{FN}$ charges of some fields should be modified as well. Indeed, if the two singlets have charges $-1$ and $-10$, for certain charges of the Yukawa couplings such as the ones we choose in this appendix, only the singlet of charge $-1$ can enter the Yukawas.}. 

In order to make this quantitative, we scanned over all possible realizations compatible with the assumptions listed at the beginning of this appendix, and we found that the least constrained vector-like models must have $v_2\geq4.5\times 10^{13}$ GeV for the unification to happen within the perturbative regime, more than one order of magnitude above the chiral model. A realization of this is obtained as follows: choose $h_u+h_d=17$, $3X_Q+X_L=-\frac{32}{3}$ and $(\phi_1,\phi_2)$ of charges $(-1,-10)$, add a set of $8$ $Q$-like and $1$ $L$-like $SU(5)$ heavy multiplets, use $6$ $Q$-like multiplets among those to generate the mass matrices and couple the rest in a chiral way to $\phi_1$ and $\phi_2$, distributing the remaining available superfields as in Table \ref{realizationVectorLikeMinimal}. 
\begin{table}[h]
\centering
$\begin{array}{|c|c|c|c|}
\hline
\text{Heavy superfield type}&\text{Number of (chiral) spectators}&\text{coupled to }\phi_1&\text{coupled to }\phi_2\\
\hline
Q&2&1&1\\
U&2&2&0\\
D&1&0&1\\
E&2&2&0\\
L&1&1&0\\
\hline
\end{array}$
\caption{Spectators chiral couplings in the minimal vector-like counterpart to model B}
\label{realizationVectorLikeMinimal}
\end{table}
This way, $A_1=A_{1,\text{SM}}+A_{1,\text{spect.}},A_2$ and $A_3$ all vanish, and one can choose $X_Q$ so that $A'_1$ vanishes as well.

\end{appendices}

\bibliographystyle{utphys}         
\bibliography{ChiralFN.bib}

\end{document}